\documentclass[11pt]{article}
\usepackage{amsmath,amsthm,bm, amscd}
\usepackage{fullpage}
\usepackage[nofullpage,full,titlepage,nousetoc,nouselot,nouselof,hylinks,final]{boaz}

\usepackage{graphicx}

\newcommand{\supp}{\mathsf{supp}}

\newcommand{\zo}{\bits}
\newcommand{\adv}{\calA}

\newcommand{\lrmac}{\mathsf{lrMAC}}
\newcommand{\lamac}{\mathsf{laMAC}}
\def\calA{{\mathcal A}}

\def\calX{{\mathcal X}}
\def\calY{{\mathcal Y}}

\DeclareMathOperator{\nm}{\nmExt} 
\DeclareMathOperator{\expect}{E}

\newcommand{\purify}{\mathsf{purify}}
\newcommand{\laext}{\mathsf{laExt}}

\theoremstyle{definition}

\newcommand{\set}[1]{\left \{{#1} \right \}}

\newcommand{\eps}{\epsilon}

\newcommand{\Cond}{\mathsf{Cond}}

\newcommand{\nmC}{\mathsf{nmCond}}

\newcommand{\Supp}{\mathsf{Supp}}
\newcommand{\Raz}{\mathsf{Raz}}

\newcommand{\Ext}{\mathsf{Ext}}
\newcommand{\Edit}{\mathsf{Edit}}

\newcommand{\nmExt}{\mathsf{nmExt}}
\newcommand{\mac}{\mathsf{MAC}}

\newcommand{\EditDis}{\mathsf{EditDis}}
\newcommand{\TExt}{\mathsf{TExt}}
\newcommand{\zuc}{\Cond}
\newcommand{\scond}{\mathsf {Scond}}

\newcommand{\BI}{\begin{itemize}}
\newcommand{\EI}{\end{itemize}}
\newcommand{\BE}{\begin{enumerate}}
\newcommand{\EE}{\end{enumerate}}

\newtheorem{thm}{Theorem}      
\newcommand{\BT}{\begin{theorem}}   \newcommand{\ET}{\end{theorem}}
\newcommand{\BD}{\begin{definition}}   \newcommand{\ED}{\end{definition}}
\newcommand{\BCR}{\begin{corollary}} \newcommand{\ECR}{\end{corollary}}
\newtheorem{constr}[thm]{Construction}
\newcommand{\BCT}{\begin{constr}} \newcommand{\ECT}{\end{constr}}
\newcommand{\BL}{\begin{lemma}}   \newcommand{\EL}{\end{lemma}}

\newcommand{\BP}{\begin{proposition}}   \newcommand{\EP}{\end{proposition}}
\newcommand{\BCM}{\begin{claim}}   \newcommand{\ECM}{\end{claim}}
\newcommand{\BF}{\begin{fact}}   \newcommand{\EF}{\end{fact}}
\newcommand{\BA}{\begin{assumption}}   \newcommand{\EA}{\end{assumption}}

\def\eps{\varepsilon}

\def\le{\leqslant} \def\ge{\geqslant}

\makeatletter
\def\ExtendSymbol#1#2#3#4#5{\ext@arrow 0099{\arrowfill@#1#2#3}{#4}{#5}}
\def\RightExtendSymbol#1#2#3#4#5{\ext@arrow 0359{\arrowfill@#1#2#3}{#4}{#5}}
\def\LeftExtendSymbol#1#2#3#4#5{\ext@arrow 6095{\arrowfill@#1#2#3}{#4}{#5}}
\makeatother
\newcommand\llrightarrow[2][]{\RightExtendSymbol{-}{-}{\rightarrow}{#1}{#2}}

\newcommand\llleftarrow[2][]{\RightExtendSymbol{\leftarrow}{-}{-}{#1}{#2}}

\newcommand{\hinf}{H_\infty}
\newcommand{\thinf}{\widetilde{H}_\infty}

\begin{document}

\begin{titlepage}
\def\thepage{}

\title{
Non-Malleable Condensers for Arbitrary Min-Entropy, and Almost Optimal Protocols for Privacy Amplification
}

\author{
Xin Li\thanks{ Supported by a Simons postdoctoral fellowship.}\\
Department of Computer Science\\
University of Washington\\
Seattle, WA 989105, U.S.A.\\
lixints@cs.washington.edu
}

\maketitle \thispagestyle{empty}

\begin{abstract}
Recently, the problem of privacy amplification with an active adversary has received a lot of attention. Given a shared $n$-bit weak random source $X$ with min-entropy $k$ and a security parameter $s$, the main goal is to construct an explicit 2-round privacy amplification protocol that achieves entropy loss $O(s)$. Dodis and Wichs \cite{DW09} showed that optimal protocols can be achieved by constructing explicit \emph{non-malleable extractors}. However, the best known explicit non-malleable extractor only achieves $k=0.49n$ \cite{Li12b} and evidence in \cite{Li12b} suggests that constructing explicit non-malleable extractors for smaller min-entropy may be hard. In an alternative approach, Li \cite{Li12} introduced the notion of a non-malleable condenser and showed that explicit non-malleable condensers also give optimal privacy amplification protocols.

In this paper, we give the first construction of non-malleable condensers for arbitrary min-entropy. Using our construction, we obtain a 2-round privacy amplification protocol with optimal entropy loss for security parameter up to $s=\Omega(\sqrt{k})$. This is the first protocol that simultaneously achieves optimal round complexity and optimal entropy loss for arbitrary min-entropy $k$. We also generalize this result to obtain a protocol that runs in $O(s/\sqrt{k})$ rounds with optimal entropy loss, for security parameter up to $s=\Omega(k)$. This significantly improves the protocol in \cite{ckor}. Finally, we give a better non-malleable condenser for linear min-entropy, and in this case obtain a 2-round protocol with optimal entropy loss for security parameter up to $s=\Omega(k)$, which improves the entropy loss and communication complexity of the protocol in \cite{Li12b}.
\end{abstract}
\end{titlepage}

\section{Introduction}
Modern cryptographic applications rely heavily on the use of randomness. Indeed, without true randomness some basic tasks such as bit commitment and encryption would become impossible. However, most of these applications require uniform random bits, yet real world random sources are rarely uniformly distributed.\ In addition, even initially uniform secret keys could be damaged by side channel attacks of an adversary.\ Naturally, the random sources we can use become imperfect, and it is therefore important to study how to run cryptographic applications using imperfect randomness.

In this general context, Bennett, Brassard, and Robert \cite{bbr} introduced the basic cryptographic question of \emph{privacy amplification}. Consider the simple model where two parties (Alice and Bob) share an $n$-bit secret key~$X$, which is weakly random. They also share a public channel which is monitored by an adversary Eve, and have access to local (non-shared) uniform private random bits. The goal now is for Alice and Bob to communicate over the channel to transform $X$ into a nearly uniform secret key, so that Eve has negligible information about it.\ To measure the randomness in $X$, we use the standard min-entropy.

\begin{definition}
The \emph{min-entropy} of a random variable~$X$ is
\[ H_\infty(X)=\min_{x \in \supp(X)}\log_2(1/\Pr[X=x]).\]
For $X \in \zo^n$, we call $X$ an $(n,H_\infty(X))$-source, and we say $X$ has
\emph{entropy rate} $H_\infty(X)/n$.
\end{definition}

We assume the adversary Eve has unlimited computational power. If Eve is passive, then this problem can be solved by using a well-studied combinatorial object called ``strong extractor". 

\smallskip\noindent
{\bf Notation.} We let $[s]$ denote the set $\set{1,2,\ldots,s}$.
For $\ell$ a positive integer,
$U_\ell$ denotes the uniform distribution on $\zo^\ell$, and
for $S$ a set,
$U_S$ denotes the uniform distribution on $S$.
When used as a component in a vector, each $U_\ell$ or $U_S$ is assumed independent of the other components.
We say $W \approx_\eps Z$ if the random variables $W$ and $Z$ have distributions which are $\eps$-close in variation distance.

\begin{definition}\label{def:strongext}
A function $\Ext : \bits^n \times \bits^d \rightarrow \bits^m$ is  a \emph{strong $(k,\eps)$-extractor} if for every source $X$ with min-entropy $k$
and independent $Y$ which is uniform on $\zo^d$,
\[ (\Ext(X, Y), Y) \approx_\eps (U_m, Y).\]
\end{definition}

Once we have a strong extractor, we can have Alice sample a fresh random string $Y$ and send it to Bob. They then both compute $\Ext(X, Y)$. Since Eve only sees $Y$, the property of the strong extractor guarantees that the output is close to uniform even given this information.\ However, if Eve is active, then the problem becomes much harder and the above simple solution fails. In this case, there has been a lot of effort in trying to achieve optimal parameters \cite{MW97, dkrs, DW09, RW03, KR09, ckor, DLWZ11, CRS11, Li12, Li12b}. More specifically, \cite{MW97} gave the first non-trivial protocol which takes one-round and works when the entropy rate of $X$ is bigger than $2/3$. \cite{dkrs} later improved this to work for entropy rate bigger than $1/2$, yet both these results suffer from the drawback that the final secret key $R$ is significantly shorter than the min-entropy of $X$. \cite{DW09} showed that it is impossible to construct one-round protocol for entropy rate less than $1/2$. The first protocol which works for entropy rate below $1/2$ appeared in \cite{RW03}, which was simplified by \cite{KR09} and shown to run in $O(s)$ rounds and achieve entropy loss $O(s^2)$, where $s$ is the security parameter of the protocol (A protocol has security parameter~$s$ if Eve cannot predict with advantage
more than $2^{-s}$ over random.  When Eve is active, we also require that Eve cannot make Alice and Bob output different secrets and not abort
with probability more than $2^{-s}$). 

\cite{DW09} improved the number of rounds to 2 but the entropy loss remains $O(s^2)$. \cite{ckor} improved the entropy loss to $O(s)$ but the number of rounds blows up to $O(s)$.  The natural open question is therefore whether there is a 2-round protocol with entropy loss $O(s)$. In the special case where the entropy rate is bigger than $1/2$, \cite{DLWZ11, CRS11, Li12} gave 2-round protocols with entropy loss $O(s)$. For any constant $0<\delta<1$, \cite{DLWZ11} also gave a protocol for $k=\delta n$ that runs in $\poly(1/\delta)$ rounds with entropy loss $\poly(1/\delta)s=O(s)$. Recently, \cite{Li12b} gave an improved protocol for $k=\delta n$ that runs in 2 rounds and achieves optimal entropy loss $2^{\poly(1/\delta)}s=O(s)$. 

In \cite{DW09}, Dodis and Wichs introduced the notion of a ``non-malleable extractor" and showed that such an object can be used to construct 2-round privacy amplification protocols with optimal entropy loss.

\begin{definition}\footnote{Following \cite{DLWZ11}, we define worst case non-malleable extractors, which is slightly different from the original definition of average case non-malleable extractors in \cite{DW09}. However, the two definitions are essentially equivalent up to a small change of parameters.}
\label{nmdef}
A function $\nm:\bits^n \times \bits^d \to \bits^m$ is a $(k,\eps)$-non-malleable extractor if,
for any source $X$ with $\hinf(X) \geq k$ and any function $\adv:\bits^d \to \bits^d$ such that $\adv(y) \neq y$ for all~$y$,
the following holds.
When $Y$ is chosen uniformly from $\bits^d$ and independent of $X$,
\[
(\nm(X,Y),\nm(X,\adv(Y)),Y) \approx_\eps (U_m,\nm(X,\adv(Y)),Y).
\]
\end{definition}

Dodis and Wichs showed that non-malleable extractors exist when $k>2m+3\log(1/\eps) + \log d + 9$ and $d>\log(n-k+1) + 2\log (1/\eps) + 7$. However, they only constructed weaker forms of non-malleable extractors.\ The first explicit construction of non-malleable extractors appears in \cite{DLWZ11}, which works for entropy $k>n/2$. Later, various improvements appear in \cite{CRS11, Li12, DY12}. However, the entropy requirement remains $k>n/2$. Recently, Li \cite{Li12b} gave the first explicit non-malleable extractor that breaks this barrier, which works for $k=(1/2-\delta)n$ for some constant $\delta>0$. \cite{Li12b} also showed a connection between non-malleable extractors and two-source extractors, which suggests that constructing explicit non-malleable extractors for smaller entropy may be hard.

Given the above background, an alternative approach seems promising. This is the notion of a non-malleable condenser introduced in \cite{Li12}. While a non-malleable extractor requires the output to be close to uniform, a non-malleable condenser only requires the output to have enough min-entropy.


\BD \cite{Li12b} A $(k, k', \e)$ non-malleable condenser is a function $\nmC: \bits^n \times \bits^d \to \bits^m$ such that given any $(n,k)$-source $X$, an independent uniform seed $Y \in \bits^d$, and any (deterministic) function $\adv: \bits^d \to \bits^d$ such that $\forall y, \adv(y) \neq y$, we have that with probability $1-\e$ over the fixing of $Y=y$, 

\[\Pr_{z' \leftarrow \nmC(X, \adv(y))}[\nmC(X, y)|_{\nmC(X, \adv(y))=z'} \text{ is } \e-\text{close to an } (m,k') \text{ source}] \geq 1-\e.\] 
\ED  

As can be seen from the definition, a non-malleable condenser is a strict relaxation of a non-malleable extractor. In \cite{Li12}, Li showed that non-malleable condensers can also be used to construct 2-round privacy amplification protocols with optimal entropy loss. Thus one can hope to construct explicit non-malleable condensers for smaller min-entropy.

\subsection{Our results}
In this paper, we indeed succeed in the above approach. We construct explicit non-malleable condensers for essentially any min-entropy. Our first theorem is as follows.

\BT
There exists a constant $C>0$ such that for any $n, k \in \N$ and $s>0$ with $k \geq C(\log n+s)^2$, there is an explicit $(k, s, 2^{-s})$-non-malleable condenser with seed length \\ $d=O(\log n+s)^2$ and output length $m=O(\log n+s)^2$.
\ET

Combining this theorem with the protocol in \cite{Li12}, we immediately obtain a 2-round privacy amplification protocol with optimal entropy loss for any security parameter up to $\Omega(\sqrt{k})$. This is the first explicit protocol that simultaneously achieves optimal parameters in both round complexity and entropy loss, for arbitrary min-entropy.

\BT
There exists a constant $C$ such that for any $\e>0$ with $k \geq C(\log n+\log(1/\e))^2$, there exists an explicit 2-round privacy amplification protocol for $(n, k)$ sources with security parameter $\log(1/\e)$, entropy loss $O(\log n+\log (1/\e))$ and communication complexity $O(\log n+\log(1/\e))^2$.
\ET

We note that except the protocol in \cite{ckor}, all previous results that work for arbitrary min-entropy $k$ only achieve security parameter up to $s=\Omega(\sqrt{k})$ like our protocol and all of them have entropy loss $\Omega(s^2)$. In this paper, we finally manage to reduce the entropy loss to $O(s)$. Thus, for this range of security parameter, ignoring the communication complexity, we essentially obtain optimal privacy amplification protocols. 

For the special case where $k=\delta n$ for some constant $0<\delta<1$, we can do better. Here we have the following theorem.

\BT \label{thm:nmc2}
For any constant $0<\delta<1$ and $k=\delta n$ there exists a constant $C=2^{\poly(1/\delta)}$ such that given any $0<s \leq k/C$, there is an explicit $(k, s, 2^{-s})$-non-malleable condenser with seed length $d=\poly(1/\delta)(\log n+s)$ and output length $m=2^{\poly(1/\delta)}(\log n+s)$.
\ET 

Combined with the protocol in \cite{Li12}, this theorem yields:

\BT
There exists an absolute constant $C_0>1$ such that for any constant $0< \delta <1$ and $k =\delta n$ there exists a constant $C_1=2^{\poly(1/\delta)}$ such that given any $\e>0$ with $C_1\log(1/\e) \leq k$, there exists an explicit 2-round privacy amplification protocol for $(n, k)$ sources with security parameter $\log(1/\e)$, entropy loss $C_0(\log n+\log (1/\e))$ and communication complexity $\poly(1/\delta)(\log n+\log(1/\e))$.
\ET

Note that for security parameter $s$, the 2-round protocol for $k=\delta n$ in \cite{Li12b} has entropy loss $2^{\poly(1/\delta)}s$ and communication complexity $2^{\poly(1/\delta)}s$. Here, we improve the entropy loss to $C_0 s$ for an absolute constant $C_0>1$ and the communication complexity to $\poly(1/\delta)s$.

Finally, one can ask what if for arbitrary min-entropy $k$, we want to achieve security parameter bigger than $\sqrt{k}$, as in \cite{ckor}. Using our techniques combined with some techniques from \cite{ckor}, we obtain the following theorem.

\BT
There exists a constant $C>1$ such that for any $n, k \in \N$ with $k \geq \log^4 n$ and any $\e>0$ with $k \geq C(\log(1/\e))$ there exists an explicit $O((\log n+\log(1/\e))/\sqrt{k})$ round privacy amplification protocol for $(n, k)$ sources with security parameter $\log(1/\e)$, entropy loss $O(\log n+\log(1/\e))$ and communication complexity $O((\log n+\log(1/\e))\sqrt{k})$.
\ET

Thus, we can essentially achieve security parameter up to $s=\Omega(k)$ with optimal entropy loss, at the price of increasing the number of rounds to $O(s/\sqrt{k})$. Note that the protocol in \cite{ckor}, though also achieving optimal entropy loss, runs in $\Omega(s)$ rounds. Thus our protocol improves their round complexity by a $\sqrt{k}$ factor. For large $k$ this is a huge improvement, especially in practice. 

\tableref{table:result} summarizes our results compared to some previous results, assuming the security parameter is $s$.

\begin{table}[ht] 
\centering 
\begin{tabular}{|l|l|l|l|l|} 
\hline Construction &  Entropy of $W$  &  Security parameter & Rounds & Entropy loss \\ 
\hline Optimal, non-explicit & $k > \log n$ & $s \leq \Omega(k)$ & $2$ & $\Theta(s + \log n)$ \\
\hline \cite{MW97} & $k > 2n/3$ & $s=\Theta(k)$ & $1$ & $(n-k)$ \\
\hline \cite{dkrs} & $k > n/2$ & $s=\Theta(k)$ & $1$ & $(n-k)$ \\
\hline \cite{RW03, KR09} & $k \geq \polylog(n)$ & $s \leq \Omega(\sqrt{k})$ & $\Theta(s + \log n)$ & $\Theta((s + \log n)^2)$ \\ 
\hline \cite{DW09} & $k \geq \polylog(n)$ & $s \leq \Omega(\sqrt{k})$ & $2$ & $\Theta((s + \log n)^2)$ \\ 
\hline \cite{ckor} & $k \geq \polylog(n)$ & $s \leq \Omega(k)$ & $\Theta(s + \log n)$ & $\Theta(s + \log n)$ \\ 
\hline \cite{DLWZ11} & $k \geq \delta n$ & $s \leq k/\poly(1/\delta)$ & $\poly(1/\delta)$ & $\poly(1/\delta)(s + \log n)$ \\ 
\hline \cite{Li12b} & $k \geq \delta n$ & $s \leq k/2^{\poly(1/\delta)}$ & $2$ & $2^{\poly(1/\delta)}(s + \log n)$ \\ 
\hline This work & $k \geq \polylog(n)$ & $s \leq \Omega(\sqrt{k})$ & $2$ & $\Theta(s + \log n)$ \\
\hline This work & $k \geq \polylog(n)$ & $s \leq \Omega(k)$ & $\Theta((s + \log n)/\sqrt{k})$ & $\Theta(s + \log n)$ \\ 
\hline This work & $k \geq \delta n$ & $s \leq k/2^{\poly(1/\delta)}$ & $2$ & $\Theta(s + \log n)$\\  \hline
\end{tabular}
\caption{\textbf{Summary of Results on Privacy Amplification with an Active Adversary}} 
\label{table:result}
\end{table}

\section{Overview of The Constructions and Techniques}
Here we give an informal overview of our constructions and the technique used. To give a clear description, we shall be imprecise sometimes.

\subsection{Non-malleable condenser for arbitrary min-entropy}
For an $(n, k)$ source $X$, our non-malleable condenser uses a uniform seed $Y=(Y_1, Y_2)$, where $Y_2$ has a bigger size than $Y_1$, say $|Y_1|=d$ and $|Y_2| = 10d$. Consider now any function $\adv(Y)=Y'=(Y_1', Y_2')$. In the following we will use letters with prime to denote variables produced with $Y'$. Since $Y' \neq Y$, we have two cases: $Y_1 =Y_1'$ or $Y_1 \neq Y_1'$. The output of our non-malleable condenser will be $Z=\nmC(X, Y)=(V_1, V_2)$. Intuitively, $V_1$ handles the case where $Y_1 =Y_1'$ and $V_2$ handles the case where $Y_1 \neq Y_1'$. We now describe the two cases separately. 

If $Y_1 =Y_1'$, then we take a strong extractor $\Ext$ and compute $W=\Ext(X, Y_1)$. Note that $W'=W$ since $Y_1 =Y_1'$. Note $Y' \neq Y$, thus we must have $Y_2' \neq Y_2$. We now fix $Y_1$ (and $Y_1'$) and conditioned on this fixing, $W=W'$ is still (close to) uniform and now $Y_2'$ is a deterministic function of $Y_2$. At this point, we can take any non-malleable extractor $\nm$ from \cite{DLWZ11, CRS11, Li12} and compute $V_1=\nm(W, Y_2)$. Since $W$ is uniform, by the property of the non-malleable extractor we have that $V_1$ is (close to) uniform even conditioned on the fixing of $V_1'$ and $(Y_2, Y_2')$. Now let the size of $V_1$ be bigger than the size of $V_2$, say $|V_1| \geq |V_2|+s$. Thus the further conditioning on the fixing of $V_2'$ will still leave $V_1$ with entropy roughly $s$. This takes care of our first case.

If $Y_1 \neq Y_1'$, then we first fix $(Y_1, Y_1')$. Note that fixing $Y_1'$ may cause $Y_2$ to lose entropy. However, since $|Y_2| = 10|Y_1|$, conditioned on this fixing $Y_2$ still has entropy rate roughly $9/10$, and now $Y_2'$ is a deterministic function of $Y_2$. We further fix $W'$, which is now a deterministic function of $X$. As long as the entropy of $X$ is larger than the size of $W$, conditioned on this fixing $X$ still has a lot of entropy. Note that after these fixings $X$ and $Y_2$ are still independent.\ Now, we use $X$ and $Y_2$ to perform an alternating extraction protocol. Specifically, take the first $3d$ bits of $Y_2$ to be $S_0$, we compute the following random variables: $R_0=\Raz(S_0, X), S_1=\Ext(Y_2, R_0), R_1=\Ext(X, S_1), S_2=\Ext(Y_2, R_1), R_2=\Ext(X, S_2), \cdots, S_t=\Ext(Y_2, R_{t-1}), R_t=\Ext(X, S_t)$. Here $\Raz$ is the two source extractor in \cite{Raz05}, which works as long as the first source has entropy rate $>1/2$, and $\Ext$ is a strong extractor. We take $t=4d$ and let each $R_i$ output $s$ bits. Note that in the first step $S_0$ roughly has entropy rate $2/3$, thus we need to use the two-source extractor $\Raz$. In all subsequent steps $S_i, R_i$ are (close to) uniform, thus it suffices to use a strong extractor. 

In the above alternating extraction protocol, as long as the size of each $(S_i, R_i)$ is relatively small, one can show that for any $i$, $R_i$ is (close to) uniform conditioned on $\{R_j, R_j', j<i\}$ and $(Y_2, Y_2')$ (recall $\{R_j'\}$ are the random variables produced by $Y_2'$ instead of $Y_2$). Next, we borrow some ideas from \cite{DW09}. Specifically, there they showed an efficient map $f$ from a string with $d$ bits to a subset of $[4d]$, such that for any $\mu \in \bits^d$, $|f(\mu)|=2d$. Moreover, for any $\mu \neq \mu'$, there exists a $j \in [4d]$ such that $|f(\mu)^{\geq j}| > |f(\mu')^{\geq j}|$, where $f(\mu)^{\geq j}$ denotes the subset of $f(\mu)$ which contains all the elements $\geq j$. Now, let $R=(R_1, \cdots, R_t)$, we define a ``look-ahead" MAC $\lamac$ such that for any $\mu \in \bits^d$, $\lamac_R(\mu)=\{R_i\}_{i \in f(\mu)}$. Now our $V_2$ is computed as $V_2=\lamac_R(Y_1)$. Note that since we have fixed $(Y_1, Y_1')$, we can now view them as two different strings in $\bits^d$. Thus, there exists a $j \in [4d]$ such that $|f(Y_1)^{\geq j}| > |f(Y_1')^{\geq j}|$. Now let $\bar{R}$ be the concatenation of those $R_i$'s in $V_2$ with $i \geq j$ and $\bar{R}'$ be the corresponding variable for $V_2'$, then the size of $\bar{R}$ is bigger than the size of $\bar{R}'$ by at least $s$. Moreover, $\bar{R}$ is (close to) uniform conditioned on the fixing of $\{R_i', i<j\}$ and $(Y_2, Y_2')$. Thus $\bar{R}$ roughly has entropy $s$ even conditioned on the fixing of $\bar{R}', \{R_i', i<j\}$ and $(Y_2, Y_2')$, which also determines $V_2'$. Since we have fixed $W'$ before, $V_1'$ is also fixed. Thus we have that $Z$ has entropy roughly $s$ even conditioned on the fixing of $Z'$ and $(Y_2, Y_2')$. This takes care of our second case. 

Thus, we obtain a non-malleable condenser for any min-entropy. However, since in the alternating extraction protocol each $R_i$ outputs $s$ bits, and we need $d=\Omega(s)$ to achieve error $2^{-s}$, the entropy of $X$ has to be larger than $4ds=\Omega(s^2)$. Thus we can only achieve $s$ up to $\Omega(\sqrt{k})$.

\subsection{Privacy amplification protocol}
Combined with the techniques in \cite{Li12b}, our non-malleable condenser immediately gives a 2-round privacy amplification protocol with optimal entropy loss for any min-entropy, with security parameter $s$ up to $\Omega(\sqrt{k})$. To better illustrate the key idea, we also give a slightly simpler 2-round protocol with optimal entropy loss, without using the non-malleable condenser. Assuming the security parameter we want to achieve is $s$, we now describe the protocol.

In the first round, Alice samples 3 random strings $(Y_1, Y_2, Y_3)$ from her private random bits and sends them to Bob, where Bob receives $(Y_1', Y_2', Y_3')$. Assume that $|Y_1|=d, |Y_2|=10d, |Y_3|=50d$. Take a strong extractor $\Ext$ and Alice and Bob each computes $R_1=\Ext(X, Y_1)$ and $R_1'=\Ext(X, Y_1')$ respectively. $R_1, R_1'$ each has $4s$ bits. Next, Alice and Bob each uses $(X, Y_2)$ and $(X, Y_2')$ to perform the alternating extraction protocol we described above, where they compute $R_2=(R_{21}, \cdots, R_{2t})$ and $R_2'=(R_{21}', \cdots, R_{2t}')$ respectively, with $t=4d$. Using $R_2$ and $R_2'$, they compute $Z=\lamac_{R_2}(Y_1)$ and $Z'=\lamac_{R_2'}(Y_1')$ respectively. 

In the second round, Bob samples a random string $W'$ from his private random bits and sends it to Alice, where Alice receives $W$. Together with $W'$, Bob also sends two tags $(T_1', T_2')$, where Alice receives $(T_1, T_2)$. For $T_1'$, Bob takes the two-source extractor $\Raz$ and computes $T_1'=\Raz(Y_3', Z')$. Let $T_1'$ output $s$ bits. For $T_2'$, Bob takes a standard message authentication code (MAC) and computes $T_2'=\mac_{R_1'}(W')$, using $R_1'$ as the key. Bob then computes $R_B=\Ext(X, W')$ as the final output. Alice will check whether $T_1=\Raz(Y_3, Z)$ and $T_2=\mac_{R_1}(W)$. If either test fails, Alice rejects. Otherwise Alice computes $R_A=\Ext(X, W)$ as the final output.

As before, the analysis can be divided into two cases: $Y_1 =Y_1'$ and $Y_1 \neq Y_1'$. In the first case, we have $R_1=R_1'$ and is (close to) uniform and private. Thus $R_1$ can be used in the MAC to authenticate $W'$ to Alice. Although $T_1'$ may give some information about $R_1$, note that $R_1$ has size $4s$ and $T_1'$ has size $s$. Thus even conditioned on $T_1'$, $R_1$ has entropy roughly $3s$. We note that the MAC works as long as the entropy rate of $R_1$ is bigger than $1/2$. Thus in this case Bob can indeed authenticate $W'$ to Alice and they will agree on a uniform and private final output.

In the second case, again we can first fix $(Y_1, Y_1')$ and $R_1'$. As before we have that after this fixing, $Y_2$ still has entropy rate roughly $9/10$, $X$ still has a lot of entropy, and $X$ is independent of $(Y_2, Y_3)$. Now we can view $(Y_1, Y_1')$ as two different strings and by the same analysis before, $Z$ roughly has entropy $s$ conditioned on the fixing of $Z'$ and $(Y_2, Y_2')$. Note that after this fixing $Y_3$ still has entropy rate $>1/2$, and $Y_3'$ is a deterministic function of $Y_3$. Since $\Raz$ is a strong two-source extractor, we have that $\Raz(Y_3, Z)$ is (close to) uniform conditioned on $(Y_3', Z', R_1', W')$, which determines $(T_1', T_2')$. Thus, in this case Alice will reject with probability $1-2^{-s}$, since the probability that Eve guesses $\Raz(Y_3, Z)$ correctly is at most $2^{-s}$.

We note that our protocol shares some similarities with the protocol in \cite{DW09}, as they both use the alternating extraction protocol and the ``look-ahead" MAC. However, there is one important difference. The protocol in \cite{DW09} uses the look-ahead MAC to authenticate the string $W'$ that Bob sends to Alice in the second round. The look-ahead MAC has size $\Omega(s^2)$ and is revealed in the second round, which causes an entropy loss of $\Omega(s^2)$. Our protocol, on the other hand, uses the look-ahead MAC to authenticate the string $Y_1$ that Alice sends to Bob in the first round. Although in the protocol we do compute some variables that have size $\Omega(s^2)$ (namely $(Z, Z')$), they are computed locally by Alice and Bob, and are \emph{never} revealed in the protocol to Eve. Instead, what is revealed to Eve is $T_1'=\Raz(Y_3', Z')$, which only has size $O(s)$. In other words, in the case where $Y_1 \neq Y_1'$, since we know that $Z$ has entropy $s$ conditioned on $Z'$, we can apply another extractor $\Raz$ to $Z$ and $Z'$ respectively, such that the resulting variable $T_1'$ only has size $O(s)$ and $\Raz(Y_3, Z)$ is (close to) uniform conditioned on $T_1'$. This brings the entropy loss down to $O(s)$.

One might think that the same trick can also be applied to the protocol in \cite{DW09}. However, this is not the case. The reason is that conditioned on $(Y, Y')$, all the random variables in our protocol that are used to authenticate $W'$ are $(R_1, T_1, R_1', T_1')$, which are deterministic functions of $X$ and have size $O(s)$. Thus in the case where Bob successfully authenticates $W'$ to Alice, we can fix them and conditioned on the fixing, $X$ and $W$ are still independent. This results in a protocol with optimal entropy loss. In the protocol in \cite{DW09}, conditioned on $(Y, Y')$, the random variables that are used to authenticate $W'$ include the output of the look-ahead extractor, which has size $\Omega(s^2)$. Thus conditioning on this random variable will cause $X$ to lose entropy $\Omega(s^2)$. On the other hand, we cannot simply apply another extractor to this MAC to reduce the output size, since then the output will be a function of $W$ and $X$, and thus conditioned on the fixing of it $W$ and $X$ will no longer be independent.

We now describe our protocol for security parameter $s> \sqrt{k}$. The very high level strategy is as follows. At the beginning of the protocol, Alice samples a random string $Y$ from her private random bits with $d_1=O(s)$ bits and sends it to Bob, where Bob receives $Y'$. They each compute $R=\Ext(X, Y)$ and $R'=\Ext(X, Y')$ respectively, by using a strong extractor $\Ext$. At the end of the protocol, Bob samples a random string $W'$ from his private random bits with $d_1$ bits and sends it to Alice, together with a tag $T=\mac_{R'}(W')$. Alice receives $(W, T)$. Bob will compute $R_B=\Ext(X, W')$ as his final output and Alice will check if $T=\mac_R(W)$. If the test fails then Alice rejects. Otherwise she will compute $R_A=\Ext(X, W)$ as her final output. In the case where $Y=Y'$, again we will have that $R=R'$ and is uniform and private. Thus in this case Bob can authenticate  $W'$ to Alice by using a MAC and $R'$ as the key. We will now modify the protocol to ensure that if $Y \neq Y'$, then with probability $1-2^{-s}$  either Alice or Bob will reject. 

If $s < \sqrt{k}$ then we can use our 2-round protocol described above. However, we want to achieve $s > \sqrt{k}$ and $X$ does not have enough entropy for the 2-round protocol. On the other hand, we note that we can still use the 2-round protocol to send a substring of $Y$ with $s'=\Omega(\sqrt{k})$ bits to Bob, such that if Eve changes this string, then with probability $1-2^{-s'}$ Alice will reject. The key observation now is that after running this 2-round protocol, conditioned on the transcript, $X$ only loses $O(s')$ entropy. Thus $X$ still has entropy $k-O(\sqrt{k})$ in Eve's view. Therefore, we can run the 2-round protocol again, using fresh random strings sampled from Alice's private random bits. This will send another substring of $Y$ with $s'=\Omega(\sqrt{k})$ bits to Bob. As long as $X$ has enough entropy, we can keep doing this and it will take us $O(s/\sqrt{k})$ rounds to send the entire $Y$ to Bob, while the entropy loss is $O(s')O(s/\sqrt{k})=O(s)$. Thus as long as $k \geq Cs$ for a sufficiently large constant $C$, the above approach will work. 

However, the simple idea described above is not enough. The reason is that to change $Y$, Eve only needs to change one substring, and she can succeed with probability $2^{-s'}>> 2^{-s}$. To fix this, we modify the protocol to ensure that, if Eve changes $Y$ to $Y' \neq Y$, then she has to change $\Omega(s/\sqrt{k})$ substrings, i.e., a constant fraction of the substrings. This is where we borrow some ideas from \cite{ckor}. Specifically, instead of having Alice just send substrings of $Y$ to Bob, we will use an asymptotically good code for edit errors and have Alice send substrings of the encoding of $Y$ to Bob. More specifically, let $M=\Edit(Y)$ be the encoding of $Y$, which has size $O(d_1)$. At the beginning of the protocol, Alice will send $Y$ to Bob, where Bob receives $Y'$. Next, our protocol will run in $L=O(s/\sqrt{k})$ phases, with each phase consisting of two rounds. In phase $i$, Alice will send the $i$'th substring $M_i$ of $M$ to Bob, where $M_i$ has $d_2=\Theta(\sqrt{k})$ bits.  In the first round of phase $i$, Alice samples two random strings $(Y_{i2}, Y_{i3})$ from her private random bits and sends them to Bob, together with $M_i$. Bob receives $(M_i', Y_{i2}', Y_{i3}')$. We will let $|Y_{i3}| \geq 10|Y_{i2}|$. As in the previous 2-round protocol, Alice will use $X$ and $Y_{i2}$ to perform an alternating extraction protocol, where she computes $R_i=(R_{i1}, \cdots, R_{it})$ with $t=4d_2$ and $Z_i=\lrmac_{R_i}(M_i)$. Correspondingly, Bob will compute $R_i'$ and $Z_i'=\lrmac_{R_i'}(M_i')$, using $X$ and $Y_{i2}'$. In the second round, Bob will send $T_i'=\Raz(Y_{i3}', Z_i')$ to Alice, where Alice receives $T_i$. Alice will now check if $T_i = \Raz(Y_{i3}, Z_i)$ and she rejects if the test fails. By the same analysis of the 2-round protocol, if Eve changes the substring $M_i$ to $M_i' \neq M_i$, then with probability $1-2^{-\Omega(\sqrt{k})}$ Alice will reject.

To synchronize between Alice and Bob, in the second round of phase $i$, we will also have Bob sample a fresh random string $W_i'$ and send it as a challenge to Alice, together with $T_i'$. Alice receives $(W_i, T_i)$. Now if Alice does not reject, then she will also compute a response $V_i=\Ext(X, W_i)$ and send it back to Bob in the first round of phase $i+1$. Bob will receive $V_i'$  and then check if $V_i'=\Ext(X, W_i')$. If the test fails then he rejects. Otherwise he proceeds as before. At the end of the protocol, Bob will first check if the received codeword $M'=M_1' \circ \cdots \circ M_L'$ is indeed equal to $\Edit(Y')$. If the test fails he rejects. Otherwise he proceeds as before. This gives our whole protocol. 

For the analysis, by the property of the code, if Eve wants to change $M=\Edit(Y)$ to $M'=\Edit(Y')$ with $Y' \neq Y$, then she has to make $\Omega(d_1)$ edit operations (insertion, deletion or altering). Since changing one substring costs at most $\sqrt{k}$ edit operations, Eve has to change at least $\Omega(s/\sqrt{k})$ substrings. We then show that as long as $X$ has an extra entropy of $O(s)$, for a constant fraction of these changes, conditioned on the event that Eve has successfully made all previous changes, the probability that Eve can make this change successfully is at most $2^{-\Omega(\sqrt{k})}$. Thus the overall probability that Eve can change $M$ to $M'$ without causing either Alice or Bob to reject is at most $(2^{-\Omega(\sqrt{k})})^{\Omega(s/\sqrt{k})}=2^{-\Omega(s)}$. The round complexity is $O(s/\sqrt{k})$ and the communication complexity is $O(s\sqrt{k})$ since in each phase, the communication complexity is $O(k)$. 

\subsection{Non-malleable condenser for linear min-entropy}
Our non-malleable condenser for linear min-entropy is similar to the construction for arbitrary min-entropy, except we use a different alternating extraction protocol, namely that in \cite{Li12b}. Specifically, we will again use a seed $Y=(Y_1, Y_2)$, where $|Y_1|=d$ and $|Y_2| \geq 10d$. The output will also be $Z=(V_1, V_2)$. For any function $\adv(Y)=Y'=(Y_1', Y_2')$, we still have two cases: $Y_1 =Y_1'$ or $Y_1 \neq Y_1'$. 

If $Y_1 =Y_1'$, then again we take a strong extractor $\Ext$ and compute $W=\Ext(X, Y_1)$ and $V_1=\nm(W, Y_2)$. As long as $|V_1| \geq |V_2|+s$, this takes care of our first case. 

If $Y_1 \neq Y_1'$, then again we first fix $(Y_1, Y_1')$ and $W'$. Conditioned on this fixing $Y_2$ still has entropy rate roughly $9/10$, and now $Y_2'$ is a deterministic function of $Y_2$. Moreover $X$ still has a lot of entropy and is independent of $Y_2$. Now we use the alternating extraction protocol in \cite{Li12b}. More specifically, since $X$ has min-entropy $k=\delta n$ we can apply a somewhere condenser in \cite{BarakKSSW05, Raz05, Zuc07} to $X$ and obtain $\bar{X}=(X_1, \cdots, X_C)$ with $C=\poly(1/\delta)$ such that at least one $X_i$ has entropy rate $0.9$. In \cite{Li12b}, Li showed that as long as $k \geq 2^{\poly(1/\delta)}s$, one can use $X, \bar{X}, Y_1, Y_2$ to perform an alternating extraction protocol and obtain $V_2$ with size $2^{\poly(1/\delta)}s$, such that when $Y_1 \neq Y_1'$, $V_2$ roughly has entropy $s$ conditioned on the fixing of $V_2'$ and $(Y_2, Y_2')$. Since we have fixed $(Y_1, Y_1')$ and $W'$ before, this means that $Z$ roughly has entropy $s$ conditioned on the fixing of $Z'$ and $(Y, Y')$.

Combined with the protocol in \cite{Li12}, we thus reduce the entropy loss of the protocol in \cite{Li12b} to $O(s)$ for an absolute constant $O(\cdot)$ and the communication complexity to $\poly(1/\delta)s$.\\

\noindent{\bf Organization.}
After some preliminaries, we give the formal definition of the privacy amplification problem in section~\ref{sec:privacy}. We define alternating extraction in section~\ref{sec:alter}. We give our non-malleable condenser for arbitrary min-entropy in section~\ref{sec:nmc}, and the general privacy amplification protocol in section~\ref{sec:protocol}. In section~\ref{sec:linear} we give our  non-malleable condenser for linear min-entropy. We conclude with some open problems in section~\ref{sec:conc}.

\section{Preliminaries} \label{sec:prelim}
We often use capital letters for random variables and corresponding small letters for their instantiations. Let $|S|$ denote the cardinality of the set~$S$.
All logarithms are to the base 2.

\subsection{Probability distributions}
\begin{definition} [statistical distance]Let $W$ and $Z$ be two distributions on
a set $S$. Their \emph{statistical distance} (variation distance) is
\begin{align*}
\Delta(W,Z) \eqdef \max_{T \subseteq S}(|W(T) - Z(T)|) = \frac{1}{2}
\sum_{s \in S}|W(s)-Z(s)|.
\end{align*}
\end{definition}

We say $W$ is $\eps$-close to $Z$, denoted $W \approx_\eps Z$, if $\Delta(W,Z) \leq \eps$.
For a distribution $D$ on a set $S$ and a function $h:S \to T$, let $h(D)$ denote the distribution on $T$ induced by choosing $x$ according to $D$ and outputting $h(x)$.

\subsection{Somewhere Random Sources, Extractors and Condensers}

\begin{definition} [Somewhere Random sources] \label{def:SR} A source $X=(X_1, \cdots, X_t)$ is $(t \times r)$
  \emph{somewhere-random} (SR-source for short) if each $X_i$ takes values in $\bits^r$ and there is an $i$ such that $X_i$ is uniformly distributed.
\end{definition}

\BD
An elementary somewhere-k-source is a 	vector of sources $(X_1, \cdots, X_t)$, such that some $X_i$ is a $k$-source. A somewhere $k$-source is a convex combination of elementary somewhere-k-sources.
\ED

\BD
A function $C: \bits^n \times \bits^d \to \bits^m$ is a $(k \to l, \e)$-condenser if for every $k$-source $X$, $C(X, U_d)$ is $\e$-close to some $l$-source. When convenient, we call $C$ a rate-$(k/n \to l/m, \e)$-condenser.   
\ED

\BD
A function $C: \bits^n \times \bits^d \to \bits^m$ is a $(k \to l, \e)$-somewhere-condenser if for every $k$-source $X$, the vector $(C(X, y)_{y \in \bits^d})$ is $\e$-close to a somewhere-$l$-source. When convenient, we call $C$ a rate-$(k/n \to l/m, \e)$-somewhere-condenser.   
\ED

\begin{definition}

A function $\TExt : \bits^{n_1} \times \bits^{n_2} \rightarrow \bits^m$ is  a \emph{strong two source extractor} for min-entropy $k_1, k_2$ and error $\e$ if for every independent  $(n_1, k_1)$ source $X$ and $(n_2, k_2)$ source $Y$, 

\[ |(\TExt(X, Y), X)-(U_m, X)| < \e\]

and

\[ |(\TExt(X, Y), Y)-(U_m, Y)| < \e,\]
where $U_m$ is the uniform distribution on $m$ bits independent of $(X, Y)$. 
\end{definition}




\subsection{Average conditional min-entropy}
\label{avgcase}

Dodis and Wichs originally defined non-malleable extractors with respect to average conditional min-entropy, a notion defined by
Dodis, Ostrovsky, Reyzin, and Smith \cite{dors}.

\begin{definition}
The \emph{average conditional min-entropy} is defined as
\[ \thinf(X|W)= - \log \left (\expect_{w \leftarrow W} \left [ \max_x \Pr[X=x|W=w] \right ] \right )
= - \log \left (\expect_{w \leftarrow W} \left [2^{-\hinf(X|W=w)} \right ] \right ).
\]
\end{definition}

Average conditional min-entropy tends to be useful for cryptographic applications.
By taking $W$ to be the empty string, we see that average conditional min-entropy is at least as strong as min-entropy.
In fact, the two are essentially equivalent, up to a small loss in parameters.

We have the following lemmas.

\begin{lemma} [\cite{dors}]
\label{entropies}
For any $s > 0$,
$\Pr_{w \leftarrow W} [\hinf(X|W=w) \geq \thinf(X|W) - s] \geq 1-2^{-s}$.
\end{lemma}

\BL [\cite{dors}] \label{lem:amentropy}
If a random variable $B$ has at most $2^{\ell}$ possible values, then $\thinf(A|B) \geq \hinf(A)-\ell$.
\EL

To clarify which notion of min-entropy and non-malleable extractor we mean, we use the term \emph{worst-case non-malleable extractor} when we refer to
our Definition~\ref{nmdef}, which is with respect to traditional (worst-case) min-entropy, and \emph{average-case non-malleable extractor} to refer to
he original definition of Dodis and Wichs, which is with respect to average conditional min-entropy.

\begin{corollary}
A $(k,\eps)$-average-case non-malleable extractor is a $(k,\eps)$-worst-case non-malleable extractor.
For any $s>0$, a $(k,\eps)$-worst-case non-malleable extractor is a $(k+s,\eps + 2^{-s})$-average-case non-malleable extractor.
\end{corollary}

Throughout the rest of our paper, when we say non-malleable extractor, we refer to the worst-case non-malleable extractor of Definition~\ref{nmdef}.

\subsection{Prerequisites from previous work}
One-time message authentication codes (MACs) use a shared random key to authenticate a message in the information-theoretic setting.
\begin{definition} \label{def:mac}
A function family $\{\mac_R : \bits^{d} \to \bits^{v} \}$ is a $\e$-secure one-time MAC for messages of length $d$ with tags of length $v$ if for any $w \in \bits^{d}$ and any function (adversary) $A : \bits^{v} \to \bits^{d} \times \bits^{v}$,

\[\Pr_R[\mac_R(W')=T' \wedge W' \neq w \mid (W', T')=A(\mac_R(w))] \leq \e,\]
where $R$ is the uniform distribution over the key space $\bits^{\ell}$.
\end{definition}

\begin{theorem} [\cite{KR09}] \label{thm:mac}
For any message length $d$ and tag length $v$,
there exists an efficient family of $(\lceil  \frac{d}{v} \rceil 2^{-v})$-secure
$\mac$s with key length $\ell=2v$. In particular, this $\mac$ is $\eps$-secure when
$v = \log d + \log (1/\e)$.\\
More generally, this $\mac$ also enjoys the following security guarantee, even if Eve has partial information $E$ about its key $R$.
Let $(R, E)$ be any joint distribution.
Then, for all attackers $A_1$ and $A_2$,

\begin{align*}
\Pr_{(R, E)} [&\mac_R(W')=T' \wedge W' \neq W \mid W = A_1(E), \\ &~(W', T') = A_2(\mac_R(W), E)] \leq \left \lceil  \frac{d}{v} \right \rceil 2^{v-\thinf(R|E)}.
\end{align*}

(In the special case when $R\equiv U_{2v}$ and independent of $E$, we get the original bound.)
\end{theorem}

\begin{remark}
Note that the above theorem indicates that the MAC works even if the key $R$ has average conditional min-entropy rate $>1/2$.
\end{remark}

Sometimes it is convenient to talk about average case seeded extractors, where the source $X$ has average conditional min-entropy $\thinf(X|Z) \geq k$ and the output of the extractor should be uniform given $Z$ as well. The following lemma is proved in \cite{dors}.

\BL \cite{dors}
For any $\delta>0$, if $\Ext$ is a $(k, \e)$ extractor then it is also a $(k+\log(1/\delta), \e+\delta)$ average case extractor.
\EL

\BT [\cite{BarakKSSW05, Raz05, Zuc07}] \label{thm:swcondenser}
For any constant $\beta, \delta>0$, there is an efficient family of rate-$(\delta \to 1-\beta, \e=2^{-\Omega(n)})$-somewhere condensers $\zuc: \bits^n \to (\bits^m)^D$ where $D=O(1)$ and $m=\Omega(n)$. 

\ET

For a strong seeded extractor with optimal parameters, we use the following extractor constructed in \cite{GuruswamiUV09}.

\BT [\cite{GuruswamiUV09}] \label{thm:optext} 
For every constant $\alpha>0$, and all positive integers $n,k$ and any $\e>0$, there is an explicit construction of a strong $(k,\e)$-extractor $\Ext: \bits^n \times \bits^d \to \bits^m$ with $d=O(\log n +\log (1/\e))$ and $m \geq (1-\alpha) k$. It is also a strong $(k, \e)$ average case extractor with $m \geq (1-\alpha) k-O(\log n+\log (1/\e))$.
\ET

We need the following construction of strong two-source extractors in \cite{Raz05}.
\begin{theorem}[\cite{Raz05}] \label{thm:razweakseed} For any
  $n_1,n_2,k_1,k_2,m$ and any $0 < \delta < 1/2$ with

\begin{itemize}
\item $n_1 \geq 6 \log n_1 + 2 \log n_2$ \item $k_1 \geq (0.5 +
\delta)n_1 + 3 \log n_1 + \log n_2$ \item $k_2 \geq 5 \log(n_1 -
k_1)$ \item $m \leq \delta \min[n_1/8,k_2/40] - 1$
\end{itemize}

There is a polynomial time computable strong 2-source extractor
$\Raz : \bits^{n_1} \times \bits^{n_2} \rightarrow \bits^m$ for
min-entropy $k_1, k_2$ with error $2^{-1.5 m}$.
\end{theorem}

\BT \cite{DLWZ11, CRS11, Li12} \label{thm:enmext}
For every constant $\delta>0$, there exists a constant $\beta>0$ such that for every $n, k \in \N$ with $k \geq (1/2+\delta)n$ and $\e>2^{-\beta n}$ there exists an explicit $(k, \e)$ non-malleable extractor with seed length $d=O(\log n+\log \e^{-1})$ and output length $m=\Omega(n)$.
\ET

The following standard lemma about conditional min-entropy is implicit in \cite{NisanZ96} and explicit in \cite{MW97}.

\begin{lemma}[\cite{MW97}] \label{lem:condition} 
Let $X$ and $Y$ be random variables and let ${\cal Y}$ denote the range of $Y$. Then for all $\e>0$, one has
\[\Pr_Y \left [ H_{\infty}(X|Y=y) \geq H_{\infty}(X)-\log|{\cal Y}|-\log \left( \frac{1}{\e} \right )\right ] \geq 1-\e.\]
\end{lemma}

We also need the following lemma.

\BL \label{lem:jerror}
Let $(X, Y)$ be a joint distribution such that $X$ has range $\calX$ and $Y$ has range $\calY$. Assume that there is another random variable $X'$ with the same range as $X$ such that $|X-X'| = \e$. Then there exists a joint distribution $(X', Y)$ such that $|(X, Y)-(X', Y)| = \e$
\EL

\begin{proof}
First let $(X'', Y)$ be the same probability distribution as $(X, Y)$. For any $x \in \calX$, let $p''_x=\Pr[X''=x]$ and $p'_x=\Pr[X'=x]$. For any $y \in \calY$, let $p_y=\Pr[Y=y]$. Let $p''_{xy}=\Pr[X''=x, Y=y]$. Let $W=\{x \in \calX: p''_x>p'_x\}$ and $V=\{x \in \calX: p''_x<p'_x\}$. Thus we have that $\sum_{x \in W} |p''_x-p'_x| = \sum_{x \in V} |p''_x-p'_x| = \e$.

We now gradually change the probability distribution $X''$ into $X'$, while keeping the distribution $Y$ the same, as follows. While $W$ is not empty or $V$ is not empty, do the following. 

\begin{enumerate}
\item Pick $x \in W \cup V$ such that $|p''_x-p'_x|=min\{|p''_x-p'_x|, x \in W \cup V\}$. 

\item If $x \in W$, we decrease $\Pr[X''=x]$ to $p'_x$. Let $\tau=p''_x-p'_x$. To ensure this is still a probability distribution, we also pick any $\bar{x} \in V$ and increase $\Pr[X''={\bar{x}}]$ to $\Pr[X''={\bar{x}}]+ \tau$. To do this, we pick the elements $y \in \calY$ one by one in an arbitrary order and while $\tau>0$, do the following. Let $\tau'=min(p''_{xy}, \tau)$, $\Pr[X''=x, Y=y]=\Pr[X''=x, Y=y]-\tau'$,  $\Pr[X''=\bar{x}, Y=y]=\Pr[X''=\bar{x}, Y=y]+\tau'$ and $\tau=\tau-\tau'$. We then update the sets $\{p''_x\}$ and $\{p''_{xy}\}$ accordingly. Note that since $p''_x=\tau+p'_x \geq \tau$, this process will indeed end when $\tau=0$ and now $\Pr[X''=x]=p'_x$. Note that after this change we still have that $p''_{\bar{x}} \leq p'_{\bar{x}}$. Also, for any $y \in \calY$ the probability $\Pr[Y=y]$ remains unchanged. Finally, remove $x$ from $W$ and if $p''_{\bar{x}} = p'_{\bar{x}}$, remove $\bar{x}$ from $V$.

\item If $x \in V$, we increase $\Pr[X''=x]$ to $p'_x$. Let $\tau=p'_x-p''_x$. To ensure that $X''$ is still a probability distribution, we also pick any $\bar{x} \in W$ and decrease $\Pr[X''={\bar{x}}]$ to $\Pr[X''={\bar{x}}]- \tau$. To do this, we pick the elements $y \in \calY$ one by one in an arbitrary order and while $\tau>0$, do the following. Let $\tau'=min(p''_{\bar{x}y}, \tau)$, $\Pr[X''=x, Y=y]=\Pr[X''=x, Y=y]+\tau'$,  $\Pr[X''=\bar{x}, Y=y]=\Pr[X''=\bar{x}, Y=y]-\tau'$ and $\tau=\tau-\tau'$. We then update the sets $\{p''_x\}$ and $\{p''_{xy}\}$ accordingly. Note that since $p''_{\bar{x}} \geq \tau+p'_{\bar{x}}$, this process will indeed end when $\tau=0$ and we still have $p''_{\bar{x}} \geq p_{\bar{x}}$. Also, for any $y \in \calY$ the probability $\Pr[Y=y]$ remains unchanged. Finally, remove $x$ from $V$ and if $p''_{\bar{x}} = p_{\bar{x}}$, remove $\bar{x}$ from $W$.
\end{enumerate}

Note that in each iteration, at least one element will be removed from $W \cup V$. Thus the iteration will end after finite steps. When it ends, we have that $\forall x, \Pr[x''=x]=p'_x$, thus $X''=X'$. Since in each step the probability $\Pr[Y=y]$ remains unchanged, the distribution $Y$ remains the same. Finally, it is clear from the algorithm that $|(X'', Y)-(X, Y)| =\e$.
\end{proof}

Next we have the following lemma.

\BL \label{lem:econdition}
Let $X$ and $Y$ be random variables and let ${\cal Y}$ denote the range of $Y$. Assume that $X$ is $\e$-close to having min-entropy $k$. Then for any $\e'>0$

\[\Pr_Y \left [ (X|Y=y) \text{ is } \e'  \text{-close to a source with min-entropy } k-\log|{\cal Y}|-\log \left( \frac{1}{\e'} \right )\right ] \geq 1-\e'-\frac{\e}{\e'}.\]
\EL

\begin{proof}
Let $\calX$ denote the range of $X$. Assume that $X'$ is a distribution on $\calX$ with min-entropy $k$ such that $|X-X'| \leq \e$. Then by lemma~\ref{lem:jerror}, there exists a joint distribution $(X', Y)$ such that 

\[|(X, Y)-(X', Y)| \leq \e.\]

Now for any $y \in \calY$, let $\Delta_y = \sum_{x \in \calX}|\Pr[X=x, Y=y]-\Pr[X'=x, Y=y]|$. Then we have 

\[\sum_{y \in \calY} \Delta_y \leq \e. \]

For any $y \in \calY$, the statistical distance between $X|Y=y$ and $X'|Y=y$ is 

\begin{align*}
\delta_y &=\sum_{x \in \calX}|\Pr[X=x|Y=y]-\Pr[X'=x|Y=y]|\\ &=(\sum_{x \in \calX}|\Pr[X=x, Y=y]-\Pr[X'=x, Y=y]|)/(\Pr[Y=y])=\Delta_y/\Pr[Y=y].
\end{align*}

Thus if $\delta_y \geq \e'$ then $\Delta_y \geq \e' \Pr[Y=y]$. Let $B_Y = \{y: \delta_y \geq \e'\}$ then we have

\[\e'\Pr[y \in B_Y]=\sum_{y \in B_Y}\e'\Pr[Y=y] \leq \sum_{y \in B_Y} \Delta_y \leq \sum_{y \in \calY} \Delta_y \leq \e.\]

Thus $\Pr[y \in B_Y] \leq \frac{\e}{\e'}$. Note that when $y \notin B_y$ we have $|X|Y=y-X'|Y=y| < \e'$. Thus by \lemmaref{lem:condition} we have the statement of the lemma.
\end{proof}

\section{Privacy Amplification with an Active Adversary}\label{sec:privacy}
In this section we formally define the privacy amplification problem. We will follow \cite{DLWZ11} and define a privacy amplification protocol $(P_A, P_B)$. The protocol is executed by two parties Alice and Bob, who share a secret $X\in \bits^n$. An active, computationally unbounded adversary Eve might have some partial information $E$ about $X$ satisfying $\thinf(X|E)\ge k$. Since Eve is unbounded, we can assume without loss of generality that she is deterministic.

We assume that Eve has full control of the communication channel between the two parties. This means that Eve can arbitrarily insert, delete, reorder or modify messages sent by Alice and Bob to each other. In particular, Eve's strategy $P_E$ defines two correlated executions $(P_A,P_E)$ and $(P_E,P_B)$ between Alice and Eve, and Eve and Bob, called ``left execution'' and ``right execution'', respectively. Alice and Bob are assumed to have fresh, private and independent random bits $Y$ and $W$, respectively. $Y$ and $W$ are not known to Eve. In the protocol we use $\perp$ as a special symbol to indicate rejection. At the end of the left execution $(P_A(X,Y),P_E(E))$, Alice outputs a key $R_A\in \bits^m \cup \{\perp\}$. Similarly, Bob outputs a key $R_B \in \bits^m \cup \{\perp\}$ at the end of the right execution $(P_E(E),P_B(X,W))$. We let $E'$ denote the final view of Eve, which includes $E$ and the communication transcripts of both executions $(P_A(X,Y),P_E(E))$ and $(P_E(E),P_B(X,W)$. We can now define the security of $(P_A,P_B)$. 

\BD An interactive protocol $(P_A, P_B)$, executed by Alice and Bob on a communication channel fully controlled by an active adversary Eve, is a $(k, m, \e)$-\emph{privacy amplification protocol} if it satisfies the following properties whenever $\thinf(X|E) \geq k$:
\begin{enumerate}
\item \underline{Correctness.} If Eve is passive, then $\Pr[R_A=R_B \land~ R_A\neq \perp \land~ R_B\neq \perp]=1$.
\item \underline{Robustness.} We start by defining the notion of {\em pre-application} robustness, which states that even if Eve is active, $\Pr[R_A \neq R_B \land~ R_A \neq \perp \land~ R_B \neq \perp]\le \e$.

The stronger notion of {\em post-application} robustness is defined similarly, except Eve is additionally given the key $R_A$ the moment she completed the left execution $(P_A,P_E)$, and the key $R_B$ the moment she completed the right execution $(P_E,P_B)$. For example, if Eve completed the left execution before the right execution, she may try to use $R_A$ to force Bob to output a different key $R_B\not\in\{R_A,\perp\}$, and vice versa.
\item \underline{Extraction.} Given a string $r\in \bits^m\cup \{\perp\}$, let $\purify(r)$ be $\perp$ if $r=\perp$, and otherwise replace $r\neq \perp$ by a fresh $m$-bit random string $U_m$:  $\purify(r)\leftarrow U_m$. Letting $E'$ denote Eve's view of the protocol, we require that

\begin{align*}
\Delta((R_A, E'),(\purify(R_A), E')) \leq \e
~~~~\mbox{and}~~~~ 
\Delta((R_B, E'),(\purify(R_B), E')) \leq \e
\end{align*}

Namely, whenever a party does not reject, its key looks like a fresh random string to Eve.
\end{enumerate}
The quantity $k-m$ is called the \emph{entropy loss} and the quantity $\log (1/\e)$ is called the \emph{security parameter} of the protocol.
\ED

\section{Alternating Extraction Protocol and Look Ahead Extractor}\label{sec:alter}
An important ingredient in our construction is the following alternating extraction protocol modified from that in \cite{DW09}.

\begin{figure}[htb]
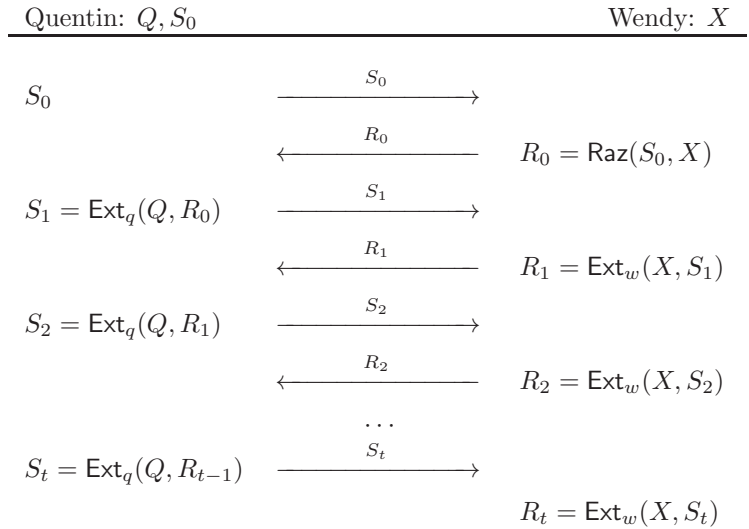

\begin{center}
\begin{small}
\begin{tabular}{l c l}
Quentin:  $Q, S_0$ & &~~~~~~~~~~Wendy: $X$ \\

\hline\\
$S_0$ & $\llrightarrow[\rule{2.5cm}{0cm}]{S_0}{} $ & \\
& $\llleftarrow[\rule{2.5cm}{0cm}]{R_0}{} $ & $R_0=\Raz(S_0, X)$ \\
$S_1=\Ext_q(Q, R_0)$ & $\llrightarrow[\rule{2.5cm}{0cm}]{S_1}{} $ & \\
& $\llleftarrow[\rule{2.5cm}{0cm}]{R_1}{} $ & $R_1=\Ext_w(X, S_1)$ \\
$S_2=\Ext_q(Q, R_1)$ & $\llrightarrow[\rule{2.5cm}{0cm}]{S_2}{} $ & \\
& $\llleftarrow[\rule{2.5cm}{0cm}]{R_2}{} $ & $R_2=\Ext_w(X, S_2)$ \\
& $\cdots$ & \\
$S_t=\Ext_q(Q, R_{t-1})$ & $\llrightarrow[\rule{2.5cm}{0cm}]{S_t}{} $ & \\
& & $R_t=\Ext_w(X, S_t)$ \\
\hline
\end{tabular}
\end{small}
\caption{\label{fig:altext}
Alternating Extraction.
}
\end{center}
\end{figure}

\textbf{Alternating Extraction.} Assume that we have two parties, Quentin and Wendy. Quentin has a source $Q$,  Wendy has a source $X$. Also assume that Quentin has a weak source $S_0$ with entropy rate $>1/2$ (which may be correlated with $Q$). Suppose that $(Q, S_0)$ is kept secret from Wendy and $X$ is kept secret from Quentin.  Let $\Ext_q$, $\Ext_w$ be strong seeded extractors with optimal parameters, such as that in \theoremref{thm:optext}. Let $\Raz$ be the strong two-source extractor in \theoremref{thm:razweakseed}. Let $d$ be an integer parameter for the protocol. For some integer parameter $t>0$, the \emph{alternating extraction protocol} is an interactive process between Quentin and Wendy that runs in $t+1$ steps. 

In the $0$'th step, Quentin sends $S_0$ to Wendy, Wendy computes $R_0=\Raz(S_0, X)$ and replies $R_0$ to Quentin, Quentin then computes $S_1=\Ext_q(Q, R_0)$. In this step $R_0, S_1$ each outputs $d$ bits. In the first step, Quentin sends $S_1$ to Wendy, Wendy computes $R_1=\Ext_w(X, S_1)$. She sends $R_1$ to Quentin and Quentin computes $S_2=\Ext_q(Q, R_1)$. In this step $R_1, S_2$ each outputs $d$ bits. In each subsequent step $i$, Quentin sends $S_i$ to Wendy, Wendy computes $R_i=\Ext_w(X, S_i)$. She replies $R_i$ to Quentin and Quentin computes $S_{i+1}=\Ext_q(Q, R_i)$. In step $i$, $R_i, S_{i+1}$ each outputs $d$ bits. Therefore, this process produces the following sequence: 

\begin{align*}
S_0, R_0=\Raz(S_0, X), & S_1=\Ext_q(Q, R_0), R_1=\Ext_w(X, S_1), \cdots, \\ &S_t=\Ext_q(Q, R_{t-1}), R_t=\Ext_w(X, S_t).
\end{align*}

\textbf{Look-Ahead Extractor.} Now we can define our look-ahead extractor. Let $Y=(Q, S_0)$ be a seed, the look-ahead extractor is defined as 

\[\laext(X,  Y)=\laext(X, (Q, S_0)) \eqdef R_1, \cdots, R_t.\]

Note that the look-ahead extractor can be computed by each party (Alice or Bob) alone in our final protocol. We now have the following lemma.

\BL \label{lem:altext}
In the alternating extraction protocol, assume that $X$ has $n$ bits and $Q$ has at most $n$ bits. Let $\e>0$ be a parameter and $d=O(\log n+\log(1/\e))> \log(1/\e)$ be the number of random bits needed in \theoremref{thm:optext} to achieve error $\e$. Assume that $X$ has min-entropy at least $12 d^2$, $Q$ has min-entropy at least $11 d^2$ and $S_0$ is a $(40d, 38d)$ source. Let $\Ext_w$ and $\Ext_q$ be strong extractors in \theoremref{thm:optext} that use $d$ bits to extract $d$ bits. Let $t=4d$.

Let $(Q', S_0')$ be another distribution on the same support of $(Q, S_0)$ such that $(Q, S_0, Q', S_0')$ is independent of $X$. Now run the alternating extraction protocol with $X$ and $(Q', S_0')$ where in each step we obtain $S_i', R_i'$.  For any $i, 0 \leq i \leq t-1$, let $\bar{S_i}=(S_0, \cdots, S_i)$, $\bar{S'_i}=(S'_0, \cdots, S'_i)$, $\bar{R_i}=(R_0, \cdots, R_i)$ and $\bar{R'_i}=(R'_0, \cdots, R'_i)$. Then for any $i, 0 \leq i \leq t-1$, we have 	
\[(R_i, \bar{S_{i-1}}, \bar{S'_{i-1}}, \bar{R_{i-1}}, \bar{R'_{i-1}}, S_i, S'_i, Q, Q') \approx_{(2i+2)\e} (U_d, \bar{S_{i-1}}, \bar{S'_{i-1}}, \bar{R_{i-1}}, \bar{R'_{i-1}}, S_i, S'_i, Q, Q').\]
\EL

\begin{proof}
We first prove the following claim.

\BCM \label{clm:condition1}
In step $0$, we have 
\[(R_0, S_0, S'_0, Q, Q') \approx_{\e} (U_d, S_0, S'_0, Q, Q')\]
and
\[(S_1, R_0, S_0, R'_0, S'_0) \approx_{3 \e} (U_d, R_0, S_0, R'_0, S'_0).\]
Moreover, conditioned on $(S_0, S'_0)$, $(R_0, R'_0)$ are both deterministic functions of $X$; conditioned on $(R_0, S_0, R'_0, S'_0)$, $(S_1, S'_1)$ are deterministic functions of $(Q, Q')$.
\ECM

\begin{proof}[Proof of the claim.] 
Note that $S_0$ is a $(40d, 38d)$ source. Thus by \theoremref{thm:razweakseed} we have that 

\[(R_0, S_0) \approx_{\e} (U_d, S_0).\]

Since conditioned on $S_0$, $R_0$ is a deterministic function of $X$, which is independent of $(Q, Q')$, we also have that 

\[(R_0, S_0, S'_0, Q, Q') \approx_{\e} (U_d, S_0, S'_0, Q, Q').\]

Now we fix $(S_0, S'_0)$ and $(R_0, R'_0)$ are both deterministic functions of $X$. Since the size of $(S_0, S_0')$ is at most $80d$, by \lemmaref{lem:condition} we have that with probability $1-\e$ over these fixings,  $Q$ is a source with entropy $10d^2$. Since $R_0, R'_0$ are both deterministic functions of $X$, they are independent of $Q$. Therefore by \theoremref{thm:optext} we have 

\[(S_1, R_0, R'_0) \approx_{\e} (U_d, R_0, R'_0).\]

Thus altogether we have that 

\[(S_1, R_0, S_0, R'_0, S'_0) \approx_{3 \e} (U_d, R_0, S_0, R'_0, S'_0)\]
Moreover, conditioned on $(R_0, S_0, R'_0, S'_0)$, $(S_1, S'_1)$ are deterministic functions of $(Q, Q')$.
\end{proof}

Now we fix $(R_0, S_0, R'_0, S'_0)$. Note that after this fixing, $S_1, S'_1$ are are deterministic functions of $(Q, Q')$. Note that with probability $1-\e$ over this fixing, $Q$ has min-entropy at least $10d^2$.

We now prove the lemma. In fact, we prove the following stronger claim.

\BCM \label{clm:condition2}
For any $i$, we have that 
\[(R_i, \bar{S_{i-1}}, \bar{S'_{i-1}}, \bar{R_{i-1}}, \bar{R'_{i-1}}, S_i, S'_i, Q, Q') \approx_{(2i+2)\e} (U_d, \bar{S_{i-1}}, \bar{S'_{i-1}}, \bar{R_{i-1}}, \bar{R'_{i-1}}, S_i, S'_i, Q, Q')\]

and

\[(S_{i+1}, \bar{S_i}, \bar{S'_i}, \bar{R_i}, \bar{R'_i}) \approx_{(2i+3)\e} (U_d, \bar{S_i}, \bar{S'_i}, \bar{R_i}, \bar{R'_i}).\]
Moreover, conditioned on $(\bar{S_{i-1}}, \bar{S'_{i-1}}, \bar{R_{i-1}}, \bar{R'_{i-1}}, S_i, S'_i)$, $(R_i, R'_i)$ are both deterministic functions of $X$; conditioned on $(\bar{S_i}, \bar{S'_i}, \bar{R_i}, \bar{R'_i})$, $(S_{i+1}, S'_{i+1})$ are deterministic functions of $(Q, Q')$.
\ECM

We prove the claim by induction on $i$. When $i=0$, the statements are already proved in \claimref{clm:condition1}. Now we assume that the statements hold for $i=j$ and we prove them for $i=j+1$.

We first fix $(\bar{S_j}, \bar{S'_j}, \bar{R_j}, \bar{R'_j})$. Since now $(S_{j+1}, S'_{j+1})$ are deterministic functions of $(Q, Q')$, they are independent of $X$. Moreover $S_{j+1}$ is $(2j+3)\e$-close to uniform. Note that the average conditional min-entropy of $X$ is at least $12d^2-2d \cdot 4d \geq 4d^2$. Therefore by \theoremref{thm:optext} we have that

\[(R_{j+1}, \bar{S_j}, \bar{S'_j}, \bar{R_j}, \bar{R'_j}, S_{j+1}, S'_{j+1}) \approx_{(2j+4)\e} (U_d, \bar{S_j}, \bar{S'_j}, \bar{R_j}, \bar{R'_j}, S_{j+1}, S'_{j+1}).\] 

Since $(S_{j+1}, S'_{j+1})$ are deterministic functions of $(Q, Q')$, we also have

\[(R_{j+1}, \bar{S_j}, \bar{S'_j}, \bar{R_j}, \bar{R'_j}, S_{j+1}, S'_{j+1}, Q, Q') \approx_{(2j+4)\e} (U_d, \bar{S_j}, \bar{S'_j}, \bar{R_j}, \bar{R'_j}, S_{j+1}, S'_{j+1}, Q, Q').\] 

Moreover, conditioned on $(\bar{S_j}, \bar{S'_j}, \bar{R_j}, \bar{R'_j}, S_{j+1}, S'_{j+1})$,  $(R_{j+1}, R'_{j+1})$ are both deterministic functions of $X$.

Next, since conditioned on $(\bar{S_j}, \bar{S'_j}, \bar{R_j}, \bar{R'_j}, S_{j+1}, S'_{j+1})$,  $(R_{j+1}, R'_{j+1})$ are both deterministic functions of $X$, they are independent of $(Q, Q')$. Moreover $R_{j+1}$ is $(2j+4)\e$-close to uniform. Note that the average conditional min-entropy of $Q$ is at least $10d^2-8d^2 =2d^2$. Therefore by \theoremref{thm:optext} we have that

\begin{align*}
& (S_{j+2}, \bar{S_j}, \bar{S'_j}, \bar{R_j}, \bar{R'_j}, S_{j+1}, S'_{j+1}, R_{j+1}, R'_{j+1}) \\ \approx_{(2j+5)\e} & (U_d, \bar{S_j}, \bar{S'_j}, \bar{R_j}, \bar{R'_j}, S_{j+1}, S'_{j+1}, R_{j+1}, R'_{j+1}).
\end{align*}

Namely, 

\[(S_{j+2}, \overline{S_{j+1}}, \overline{S'_{j+1}}, \overline{R_{j+1}}, \overline{R'_{j+1}}) \approx_{(2(j+1)+3)\e} (U_d, \overline{S_{j+1}}, \overline{S'_{j+1}}, \overline{R_{j+1}}, \overline{R'_{j+1}}).\] 
Moreover, conditioned on $(\overline{S_{j+1}}, \overline{S'_{j+1}}, \overline{R_{j+1}}, \overline{R'_{j+1}})$,  $(S_{j+2}, S'_{j+2})$ are deterministic functions of $(Q, Q')$.
\end{proof}

\section{Non-Malleable Condensers for Arbitrary Min-Entropy}\label{sec:nmc}
In this section we give our construction of non-malleable condensers for arbitrary min-entropy.

First, we need the following definitions and constructions from \cite{DW09}.

\BD \cite{DW09}
Given $S_1, S_2 \subseteq \{1, \cdots, t\}$, we say that the ordered pair $(S_1, S_2)$ is top-heavy if there is some integer $j$ such that $|S^{\geq j}_1| > |S^{\geq j}_2|$, where $S^{\geq j} \eqdef \{s \in S | s \geq j\}$. Note that it is possible that $(S_1, S_2)$ and $(S_2, S_1)$ are both top-heavy. For a collection $\Psi$ of sets $S_i \subseteq \{1, \cdots, t\}$, we say that $\Psi$ is pairwise top-heavy if every ordered pair $(S_i, S_j)$ of sets $S_i, S_j \in \Psi$ with $i \neq j$, is top-heavy.
\ED

Now, for any $m$-bit message $\mu=(b_1, \cdots, b_m)$, consider the following mapping of $\mu$ to a subset $S \subseteq \{1, \cdots, 4m\}$: 

\[f(\mu)=f(b_1, \cdots, b_m)=\{4i-3+b_i, 4i-b_i | i=1, \cdots, m\}\]

i.e., each bit $b_i$ decides if to include $\{4i-3, 4i\}$ (if $b_i=0$) or $\{4i-2, 4i-1\}$ (if $b_i=1$) in $S$. 

We now have the following lemma.

\BL \cite{DW09} \label{lem:laset}
The above construction gives a pairwise top-heavy collection $\Psi$ of $2^m$ sets $S \subseteq \{1, \cdots, t\}$ where $t=4m$. Furthermore, the function $f$ is an efficient mapping of $\mu \in \bits^m$ to $S_{\mu}$.
\EL

Now we have the following construction.

Let $r \in (\bits^{d})^t$ be the output of the look-ahead extractor defined above, i.e., $r=(r_1, \cdots, r_t)=\laext(X, (Q, S_0))$. Let $\Psi=\{S_1, \cdots, S_{2^m}\}$ be the pairwise top-heavy collection of sets constructed above. For any message $\mu \in \bits^m$, define the function $\lamac_r(\mu) \eqdef [r_i | i \in S_{\mu}]$, indexed by $r$. 

Now we can describe our construction of the non-malleable condenser.

\begin{algorithmsub}
{$\nmC(x, y)$}
{$\ell$--an integer parameter. $x$ --- a sample from an $(n,k)$-source with $k \geq 60 d^2$. $y$--an independent random seed with $y=(y_1, y_2)$ such that $y_1$ has size $d=O(\log n+\ell)> 5\ell$ and $y_2$ has size $12 d^2$.  }
{$z$ --- an $m$ bit string.} {
Let $\nm$ be the non-malleable extractor from \theoremref{thm:enmext}, with error $2^{-4\ell}$.

Let $\Ext$ be the strong extractor with optimal parameters from \theoremref{thm:optext}, with error $2^{-5\ell}$.

Let $\laext$ be the look-ahead extractor defined above, using $\Ext$ as $\Ext_q$ and $\Ext_s$. $\laext$ is set up to extract from $x$ using seed $(q, s_0)$ such that $q=y_2$ and $s_0$ is the string that contains the first $40d$ bits of $y_2$, and output a string $r \in (\bits^{d})^t$ with $t=4d$.

Let $\lamac_r(\mu)$ be the function defined above. 
} {alg:nmc}

\begin{enumerate}

\item Compute $w=\Ext(x, y_1)$ with output size $20d^2$ and $r=\laext(x, (q, s_0))$.

\item Output $z=(\nm(w, y_2), \lamac_r(y_1))$ such that $\nm(w, y_2)$ has size $8 d^2$. 
\end{enumerate}
\end{algorithmsub}

We can now prove the following theorem.

\BT \label{thm:nmc}
There exists a constant $C>0$ such that given any $s>0$, as long as $k \geq C(\log n+s)^2$, the above construction is a $(k, s, 2^{-s})$-non-malleable condenser with seed length $O(\log n+s)^2$ and output length  $O(\log n+s)^2$.
\ET

\begin{thmproof}
Let $\adv$ be any (deterministic) function such that $\forall y \in \Supp(Y), \adv(y) \neq y$. We will show that for most $y$, with high probability over the fixing of $\nmC(X, \adv(y))$, $\nmC(X, y)$ is still close to having min-entropy at least $\ell$. Let $Y'=\adv(Y)$. Thus $Y' \neq Y$. In the following analysis we will use letters with prime to denote the corresponding random variables produced with $Y'$ instead of $Y$. Let $V_1=\nm(W, Y_2)$ and $V_2=\lamac_R(Y_1)$. Thus $Z=(V_1, V_2)$. We have the following two cases.

\textbf{Case 1:} $Y_1 = Y'_1$. In this case, since $Y' \neq Y$, we must have that $Y_2 \neq Y'_2$. Now by \theoremref{thm:optext} we have that 

\[(W, Y_1) \approx_{2^{-{5\ell}}} (U, Y_1).\]

Therefore, we can now fix $Y_1$ (and thus $Y'_1$), and with probability $1-2^{-\ell}$ over this fixing, $W$ is $2^{-4\ell}$-close to uniform. Moreover, after this fixing $W$ is a deterministic function of $X$ and thus is independent of $Y_2$. Note also that after this fixing, $Y'_2$ is a deterministic function of $Y_2$. Thus by \theoremref{thm:enmext} we have that 

\[(V_1, V'_1, Y_2, Y'_2) \approx_{O(2^{-4\ell})} (U_{8 d^2}, V'_1, Y_2, Y'_2).\] 

Therefore, we can now further fix $Y_2$ (and thus $Y'_2$) and with probability at least $1-O(2^{-\ell})$ over this fixing, $(V_1, V'_1)$ is $2^{-3\ell}$-close to $(U_{8 d^2}, V'_1)$. Thus we can further fix $V'_1$, and with probability at least $1-2^{-\ell}$ over this fixing, $V_1$ is $2^{-2\ell}$-close to uniform. Now note that $V_1$ has size $8 d^2$ and $V'_2$ has size $2 d^2$. Thus by \lemmaref{lem:econdition}, we can further fix $V'_2$, and with probability at least $1-2 \cdot 2^{-\ell}$ over this fixing, $V_1$ is $2^{\ell}$-close to having min-entropy at least $8 d^2-2 d^2-\ell  \geq 5 d^2$.

Thus in this case we have shown that, with probability $1-O(2^{-\ell})$ over the fixing of $Y$, with probability $1-O(2^{-\ell})$ over the fixing of $Z'$, $Z$ is $2^{-\ell}$-close to having min-entropy at least $5 d^2> 5\ell^2$.

\textbf{Case 2:} $Y_1 \neq Y'_1$. In this case, we first fix $Y_1$ and $Y'_1$. Note that after this fixing, $W$ and $W'$ are now deterministic functions of $X$. We now further fix $W$ and $W'$ and after this fixing, $X$ and $Y_2$ are still independent. Since the total size of $(W, W')$ is $40 d^2$, by \lemmaref{lem:condition} we have that with probability $1-2^{-2\ell}$ over this fixing, $X$ still has min-entropy at least $60 d^2-40 d^2-2\ell > 12 d^2$. Note also that after this fixing, $Y'_2$ is a deterministic function of $Y_2$. However, since $Y'_1$ may be a function of $Y_2$, fixing $Y'_1$ may cause $Y_2$ to lose entropy. Note that $Y'_1$ only has size $d$, thus by \lemmaref{lem:condition}, with probability $1-2 \cdot 2^{-2\ell}$ over the fixing of $(Y_1, Y'_1)$, we have that $Y_2$ has min-entropy at least $12 d^2-d-2\ell > 11 d^2$ and $S_0$ has min-entropy at least $40d-d-2\ell > 38d$.

Now assume that $X$ has min-entropy at least $12 d^2$, $Y_2$ has min-entropy at least $11 d^2$ and $S_0$ has min-entropy at least $38d$. This happens with probability at least $1-O(2^{-\ell})$. For any $i, 0 \leq i \leq t-1$, let $\bar{S_i}=(S_0, \cdots, S_i)$, $\bar{S'_i}=(S'_0, \cdots, S'_i)$, $\bar{R_i}=(R_0, \cdots, R_i)$ and $\bar{R'_i}=(R'_0, \cdots, R'_i)$. Now by \lemmaref{lem:altext} (note that $Y_2=(Q, S_0)$) we have that for any $i, 0 \leq i \leq t-1$, 	
\[(R_i, \bar{S_{i-1}}, \bar{S'_{i-1}}, \bar{R_{i-1}}, \bar{R'_{i-1}}, S_i, S'_i, Y_2) \approx_{(2i+2)2^{-5\ell}} (U_d, \bar{S_{i-1}}, \bar{S'_{i-1}}, \bar{R_{i-1}}, \bar{R'_{i-1}}, S_i, S'_i, Y_2).\]

Therefore, we have that for any $i$,

\[(R_i, \bar{R_{i-1}}, \bar{R'_{i-1}}, Y_2) \approx_{(2i+2)2^{-5\ell}} (U_d,  \bar{R_{i-1}}, \bar{R'_{i-1}}, Y_2).\]

Thus, for any $i$, with probability $1-2^{-1.25\ell}$ over the fixing of $Y_2$, we have

\[(R_i, \bar{R_{i-1}}, \bar{R'_{i-1}}) \approx_{(2i+2)2^{-3.75\ell}} (U_d,  \bar{R_{i-1}}, \bar{R'_{i-1}}).\]

By the union bound, we have that with probability $1-t2^{-1.25\ell}$ over the fixing of $Y_2$, for any $i$,

\[(R_i, \bar{R_{i-1}}, \bar{R'_{i-1}}) \approx_{(2i+2)2^{-3.75\ell}} (U_d,  \bar{R_{i-1}}, \bar{R'_{i-1}}).\]

Consider a typical fixing of $Y_2$. Now note that $V_2=\lamac_R(Y_1)$ and $V'_2=\lamac_{R'}(Y'_1)$. Let the two sets in \lemmaref{lem:laset} that correspond to $Y_1$ and $Y'_1$ be $H$ and $H'$. Since $Y_1 \neq Y'_1$, by definition there exists $j \in [4d]$ such that $|H^{\geq j}| > |H'^{\geq j}|$. Let $l=|H^{\geq j}|$. Thus $l \leq t$ and $|H'^{\geq j}| \leq l-1$. Let $R_H$ be the concatenation of $\{R_i, i \in H^{\geq j}\}$ and $R'_{H'}$ be the concatenation of $\{R'_i, i \in H'^{\geq j}\}$.

By the above equation and the hybrid argument we have that 

\[(R_H, \bar{R_{j-1}}, \bar{R'_{j-1}}) \approx_{3t^2 \cdot 2^{-3.75\ell}} (U_{ld}, \bar{R_{j-1}}, \bar{R'_{j-1}}).\]

Thus now we can first fix $\bar{R'_{j-1}}$, and with probability $1-2^{-1.25\ell}$ over this fixing, we have

\[R_H \approx_{3t^2 \cdot 2^{-2.5\ell}} U_{ld}.\]

We now fix $R'_{H'}$. Since $|H'^{\geq j}| \leq l-1$, the size of $R'_{H'}$ is at most $(l-1)d$. Thus by \lemmaref{lem:econdition} we have that with probability at least $1-(3t^2+1) \cdot 2^{-1.25\ell}$ over this fixing, $R_H$ is $2^{-1.25\ell}$-close to having min-entropy $d-1.25\ell > \ell$. Note that after we fix $\bar{R'_{j-1}}$ and $R'_{H'}$, we have also fixed $V'_2$. Since $W'$ and $Y'_2$ are already fixed, $V'_1$ is also fixed. Thus $Z'$ is fixed. Therefore altogether we have that with probability $1-2 \cdot 2^{-2\ell}-t2^{-1.25\ell}=1-O(2^{-\ell})$ over the fixings of $Y$, with probability $1-2^{-1.25\ell}-(3t^2+1) \cdot 2^{-1.25\ell}=1-O(2^{-\ell})$ over the fixings of $Z'$, $Z$ is $2^{-1.25\ell}$-close to having min-entropy $\ell$.

Combining \textbf{Case 1} and \textbf{Case 2}, and notice that the fraction of ``bad seeds" that an adversary can achieve is at most the sum of the fraction of bad seeds in both cases. Thus by choosing an appropriate $\ell=O(s)$ we have that the construction is a $(k, s, 2^{-s})$-non-malleable condenser with seed length $O(\log n+s)^2$. 
\end{thmproof}

The following theorem is proved in \cite{Li12}.

\BT \cite{Li12}\label{thm:nmcext}
There exists a constant $C>1$ such that the following holds. For any integers $n, k$ and $\e>0$, assume that there is an explicit $(k, k', \e)$-non-malleable condenser with seed length $d$ such that $k' \geq C(\log n+\log(1/\e))$. Then there exists an explicit 2-round privacy amplification protocol for $(n, k)$ sources with entropy loss $O(\log n+\log (1/\e))$ and communication complexity $O(d+\log n+\log(1/\e))$.
\ET

Combining the above theorem and theorem~\ref{thm:nmc}, we immediately get a 2-round privacy amplification protocol with optimal entropy loss for any $(n, k)$ source. 

\BT
There exists a constant $C$ such that for any $\e>0$ with $k \geq C(\log n+\log(1/\e))^2$, there exists an explicit 2-round privacy amplification protocol for $(n, k)$ sources with security parameter $\log(1/\e)$, entropy loss $O(\log n+\log (1/\e))$ and communication complexity $O(\log n+\log(1/\e))^2$.
\ET

In fact, we have a slightly simpler protocol that uses the look-ahead extractor and MAC somewhat more directly, while achieving the same performance.

We assume that the shared weak random source has min-entropy $k$, and the error $\e$ we seek satisfies $\e<1/n$ and $k>C(\log n+\log(1/\e))^2$ for some constant $C>1$. For convenience, in the description below we introduce an ``auxiliary'' security parameter $s$. Eventually, we will set $s=\log(C'/\e)+O(1)=\log(1/\e)+O(1)$, so that $C'/2^s<\e$, for a sufficiently large constant $C'$ related to the number of ``bad'' events we need to account for. We need the following building blocks:

\begin{itemize}
\item Let $\Ext$ be a $(k,2^{-5s})$-extractor with optimal entropy loss and seed length $d=O(\log n+s)>202s$, from \theoremref{thm:optext}. Assume that $k \geq 15 d^2$.

\item Let $\Raz$ be the two source extractor from \theoremref{thm:razweakseed}.

\item Let $\mac$ be the (``leakage-resilient'') MAC, as in \theoremref{thm:mac}, with tag length $v = 2s$ and key length $\ell = 2v =  4s$. 

\item Let $\laext$ be the look-ahead extractor defined above, using $\Ext$ as $\Ext_q$ and $\Ext_s$. $\laext$ is set up to extract from $x$ using seed $(q, s_0)$ such that $q=y_2$ and $s_0$ is the string that contains the first $40d$ bits of $y_2$, and output a string $r \in (\bits^{d})^t$ with $t=4d$.

\item Let $\lamac_r(\mu)$ be the function defined above. 

\item In the protocol Alice will sample three random strings $Y_1, Y_2, Y_3$, with size $d$, $12d^2$ and $50d^2$ respectively.
\end{itemize}

\begin{figure}[htb]
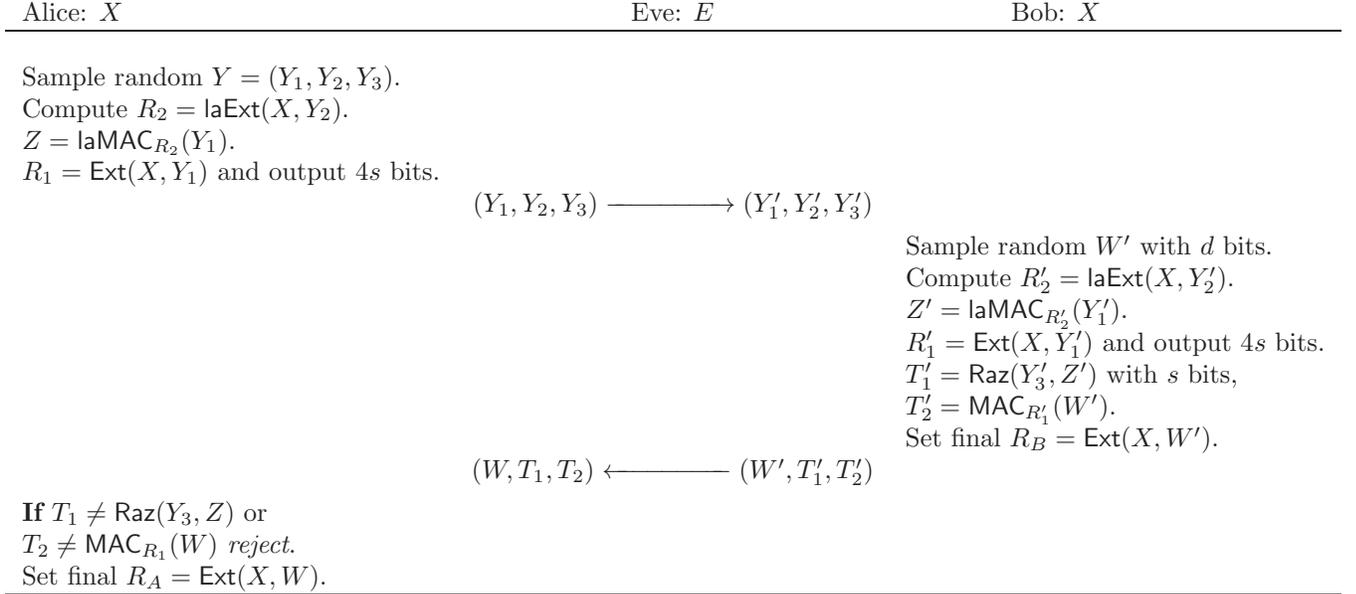

\begin{center}
\begin{small}
\begin{tabular}{l c l}
Alice:  $X$ & Eve: $E$ & ~~~~~~~~~~~~Bob: $X$ \\

\hline\\
Sample random $Y=(Y_1, Y_2, Y_3)$.& &\\
Compute $R_2=\laext(X, Y_2)$. \\ 
$Z=\lamac_{R_2}(Y_1)$. \\
 $R_1= \Ext(X, Y_1)$ and output $4s$ bits.&&\\
 & $(Y_1, Y_2, Y_3) \llrightarrow[\rule{1.5cm}{0cm}]{} (Y_1', Y_2', Y_3')$ & \\
&& Sample random $W'$ with $d$ bits.\\
&& Compute $R_2'=\laext(X, Y_2').$ \\
&& $Z'=\lamac_{R_2'}(Y_1')$. \\
&&  $R_1' = \Ext(X, Y_1')$ and output $4s$ bits.\\
&&  $T_1'= \Raz(Y'_3, Z')$ with $s$ bits, \\
&& $T_2' = \mac_{R_1'}(W')$.\\
&&  Set final $R_B = \Ext(X, W')$.\\
 & $(W,T_1, T_2) \llleftarrow[\rule{1.5cm}{0cm}]{} (W',T_1', T_2')$ & \\
{\bf If} $T_1 \neq \Raz(Y_3, Z)$ or && \\
$T_2 \neq \mac_{R_1}(W)$ {\em reject}.&&\\
Set final $R_A = \Ext(X, W)$.&&\\
\hline
\end{tabular}
\end{small}
\caption{\label{fig:AKA}
$2$-round Privacy Amplification Protocol.
}
\end{center}
\end{figure}

Using the above building blocks, the protocol is given in Figure~\ref{fig:AKA}. To emphasize the adversary Eve, we use letters with `prime' to denote all the variables seen or generated by Bob; e.g., Bob picks $W'$, but Alice may see a different $W$, etc. 

\BT \label{thm:protocol1}
Assume that $k>C(\log n+\log(1/\e))^2$ for some constant $C>1$. The above protocol is a privacy amplification protocol with security parameter $\log(1/\e)$, entropy loss $O(\log(1/\e))$ and communication complexity $O(\log(1/\e)^2)$.
\ET

\begin{proof}
The proof can be divided into two cases: whether the adversary changes $Y_1$ or not.

\textbf{Case 1:} The adversary does not change $Y_1$. In this case, note that $R_1=R_1'$ and is $2^{-5s}$-close to uniform in Eve's view (even conditioned on $Y_1, Y_2, Y_3$). Thus the property of the $\mac$ guarantees that Bob can authenticate $W'$ to Alice. However, one thing to note here is that Eve has some additional information, namely $T_1'$ which can leak information about the MAC key. On the other hand, the size of $T_1'$ is $s$, thus by \lemmaref{lem:amentropy} the average conditional min-entropy $\hinf(R_1|T_1')$ is at least $3s$. Therefore by \theoremref{thm:mac} the probability that Eve can change $W'$ to a different $W$ without causing Alice to reject is at most 

\[\left \lceil  \frac{d_1}{2s} \right \rceil 2^{2s-\thinf(R_1|T_1')}+2^{-5s} \leq O(2^{2s-3s})+2^{-5s} \leq O(2^{-s}).\]

When $W=W'$, by \theoremref{thm:optext} $R_A=R_B$ and is $2^{-5s}$-close to uniform in Eve's view. 

\textbf{Case 2:} The adversary does change $Y_1$. Thus we have $Y_1 \neq Y'_1$. Here the proof is similar to the proof of the non-malleable condenser. We first fix $Y_1$ and $Y'_1$. Note that after this fixing, $R_1$ and $R_1'$ are now deterministic functions of $X$. We now further fix $R_1$ and $R_1'$ and after this fixing, $X$ and $(Y_2, Y_3)$ are still independent. Since the total size of $(R_1, R_1')$ is $8s$, by \lemmaref{lem:condition} we have that with probability $1-2^{-2s}$ over this fixing, $X$ still has min-entropy at least $15 d^2-8s-2s > 12 d^2$. Note also that after this fixing, $Y'_2$ is a deterministic function of $(Y_2, Y_3)$. However, since $Y'_1$ may be a function of $Y_2$, fixing $Y'_1$ may cause $Y_2$ to lose entropy. Note that $Y'_1$ only has size $d$, thus by \lemmaref{lem:condition}, with probability $1-2 \cdot 2^{-2s}$ over the fixing of $(Y_1, Y'_1)$, we have that $Y_2$ has min-entropy at least $12 d^2-d-2s > 11 d^2$ and $S_0$ has min-entropy at least $40d-d-2s > 38d$.

Now assume that $X$ has min-entropy at least $12 d^2$, $Y_2$ has min-entropy at least $11 d^2$ and $S_0$ has min-entropy at least $38d$. This happens with probability at least $1-O(2^{-s})$. For any $i, 0 \leq i \leq t-1$, let $\bar{S_i}=(S_0, \cdots, S_i)$, $\bar{S'_i}=(S'_0, \cdots, S'_i)$, $\bar{R_i}=(R_0, \cdots, R_i)$ and $\bar{R'_i}=(R'_0, \cdots, R'_i)$. Again by \lemmaref{lem:altext} we have that for any $i$,

\[(R_i, \bar{S_{i-1}}, \bar{S'_{i-1}}, \bar{R_{i-1}}, \bar{R'_{i-1}}, S_i, S'_i, Y_2, Y'_2) \approx_{(2i+2)2^{-5s}} (U_d, \bar{S_{i-1}}, \bar{S'_{i-1}}, \bar{R_{i-1}}, \bar{R'_{i-1}}, S_i, S'_i, Y_2, Y'_2).\]

Thus for any $i$, we have

\[(R_i, \bar{R_{i-1}}, \bar{R'_{i-1}}, Y_2, Y'_2) \approx_{(2i+2)2^{-5s}} (U_d, \bar{R_{i-1}}, \bar{R'_{i-1}}, Y_2, Y'_2).\]

Now by the same analysis as in the proof of the non-malleable condenser (and recall that $Y_1 \neq Y'_1$), we have that with probability $1-t2^{-1.25\ell}$ over the fixing of $(Y_2, Y'_2)$, with probability at least $1-(3t^2+1) \cdot 2^{-1.25s}$ over the fixing of $Z'$, $Z$ is $2^{-1.25s}$-close to having min-entropy $d-1.25s > 200s$.

Note that we have now fixed $(Y_1, Y_1', Y_2, Y_2')$ and $(R_1, R_1', Z')$. After all these fixings, $Z$ is a deterministic function of $X$ and is $2^{-1.25s}$-close to having min-entropy $200s$. Thus $Z$ is independent of $Y_3$ (note that $Z'$ is also a deterministic function of $X$, thus fixing $Z'$ does not influence the independence of $Z$ and $Y_3$). Note that after these fixings, $Y_3'$ is a deterministic function of $Y_3$, and since the size of $(Y_1', Y_2')$ is $d+12d^2 < 13d^2$, by \lemmaref{lem:condition} $Y_3$ is $2^{-s}$-close to having min-entropy $50d^2-13d^2-s > 36d^2$. Thus by \theoremref{thm:razweakseed} we have 

\[(\Raz(Y_3, Z), Y_3, Y'_3) \approx_{O(2^{-s})} (U_s, Y_3, Y'_3).\]

Since we already fixed $(Y_1, Y_1', Y_2, Y_2')$ and $(R_1, R_1', Z')$, and $W'$ is independent of all random variables above, this also implies that 

\[(\Raz(Y_3, Z), R_1', Z', Y, Y', W') \approx_{O(2^{-s})} (U_s, R_1', Z', Y, Y', W').\]

Note that $T_1'=\Raz(Y'_3, Z')$ and $T_2' = \mac_{R_1'}(W')$. Thus we have

\[(\Raz(Y_3, Z), T_1', T_2', Y, Y', W') \approx_{O(2^{-s})} (U_s, T_1', T_2', Y, Y', W').\]

Therefore, the probability that the adversary can guess the correct $T_1$ is at most $2^{-s}+O(2^{-s})=O(2^{-s})$. For an appropriately chosen $s=\log(1/\e)+O(1)$ this is at most $\e$. Note that conditioned on the fixing of $Y$, the random variables that are used to authenticate $W'$ are $(R_1, T_1)$, which are deterministic functions of $X$ and have size $O(s)$, thus the entropy loss of the protocol is $O(\log(1/\e))$. The communication complexity can be easily verified to be $O(\log(1/\e)^2)$.
\end{proof}


\section{Improved Privacy Amplification Protocol for Smaller Error}\label{sec:protocol}
The 2-round protocol described above only works for security parameter up to $\Omega(\sqrt{k})$. In this section we generalize the above protocol and give a protocol that can achieve security parameter up to $\Omega(k)$, or equivalently, error as small as $2^{-\Omega(k)}$. First we need the following definition and theorem.

\BD For any two strings $c$ and $c'$ of length $\lambda_c$, let $\EditDis(c, c')$ denote the edit distance between $c$ and $c'$, i.e., the minimum number of single-bit insert, delete or alter operations required to change string $c$ into $c'$. 
\ED

\BD \cite{ckor} Let $m \in \bits^{\lambda_m}$. For some constant $0< e <1$, a function $\Edit: \bits^{\lambda_m} \to \bits^{\lambda_c}$ is a $(\lambda_m, e, \rho)$- code for edit errors, if $\rho \lambda_c = \lambda_m$ and the following properties are satisfied:
\begin{itemize} 
\item $c=\Edit(m)$ can be computed in polynomial (in $\lambda_m$) time, given $m$, for all $m \in \bits^{\lambda_m}$.
\item For any $m, m' \in \bits^{\lambda_m}$ with $m \neq m'$, $\EditDis(c, c') \geq e \lambda_c$, where $c=\Edit(m)$ and $c'=\Edit(m')$. 
\end{itemize}

$\rho =\frac{\lambda_m}{\lambda_c}$ is called the rate of the code.
\ED

As in \cite{ckor} the code we use is due to Schulman and Zuckerman \cite{SchulmanZ99}:

\BT  [\cite{SchulmanZ99, ckor}]\label{thm:editcode}
Let $0<e<1$ be a constant. Then for some constant $0<\rho<1$ there exists a $(\lambda_m, e, \rho)$-code for edit errors. 
\ET

We assume that the shared weak random source has min-entropy $k \geq \log^4 n$, and the error $\e$ we seek satisfies $2^{-\beta k}< \e<2^{-\Omega(\sqrt{k})}$ for some constant $\beta<1$. Again, in the description below we will introduce an ``auxiliary'' security parameter $s$ with $s=C'(\log(1/\e))$ for some sufficiently large constant $C'$. We will also use another parameter $\ell =\alpha \sqrt{k}$ for some constant $0<\alpha<1$ such that $k>C(\log n+\ell)^2$ for some constant $C>1$. We need the following building blocks:

\begin{itemize}
\item Let $\Ext_1$ be a $(k, 2^{-s})$-extractor with optimal entropy loss and seed length $d_1=O(\log n+s)=O(s)> 2s$, from \theoremref{thm:optext}. 

\item Let $\Ext_2$ be a $(k, 2^{-10\ell})$-extractor with seed length $d_2=O(\log n+\ell)=O(\ell)> 404\ell$ and output length $d_2$, from \theoremref{thm:optext}. Assume that $k \geq d_1/\rho+2s+15 d_2^2$.

\item Let $\Raz$ be the two source extractor from \theoremref{thm:razweakseed}.

\item Let $\mac$ be the (``leakage-resilient'') MAC, as in \theoremref{thm:mac}, with tag length $v = 2d_1/\rho$ and key length $2v =  4d_1/\rho$. 

\item Let $\laext$ be the look-ahead extractor defined above, using $\Ext_2$ as $\Ext_q$ and $\Ext_s$. $\laext$ is set up to extract from $x$ using seed $(q, s_0)$ such that $q=y_2$ and $s_0$ is the string that contains the first $40d_2$ bits of $y_2$, and output a string $r \in (\bits^{d_2})^t$ with $t=4d_2$.

\item Let $\lamac_r(\mu)$ be the function defined above. 

\item In each phase of the protocol Alice will sample two random strings $Y_{i2}, Y_{i3}$, with size $12d_2^2$ and $50d_2^2$ respectively.
\end{itemize}


Given these building blocks, our protocol runs in roughly $L=d_1/(\rho d_2)$ phases. The protocol is given in Figure~\ref{fig:AKA2}.

\begin{figure}[htbp]
\begin{center}
\begin{small}
\begin{tabular}{l c l}
Alice:  $X$ & Eve: $E$ & ~~~~~~~~~~~~Bob: $X$ \\

\hline\\
Sample random $Y$ with $d_1$ bits.& \fbox{{\bf Phase $1$}} &\\
Let $M=\Edit(Y)$. $M$ has length $d_1/\rho$. &&\\
Divide $M$ sequentially into $L$ blocks &&\\
$M=M_1 \circ \cdots \circ M_L$ & & \\
with each block having $d_2$ bits. && \\
Sample random $(Y_{12}, Y_{13})$.& &\\
Compute $R_1=\laext(X, Y_{12})$. \\ 
$Z_1=\lamac_{R_1}(M_1)$. \\
 & $(Y, M_1, Y_{12}, Y_{13}) \llrightarrow[\rule{0.5cm}{0cm}]{} (Y', M_1', Y_{12}', Y_{13}')$ & \\
 && Sample random $W_1'$ with $d_2$ bits.\\
&& Compute $R_1'=\laext(X, Y_{12}')$. \\
&& $Z_1'=\lamac_{R_1'}(M_1')$. \\
&&  $T_1'= \Raz(Y'_{13}, Z_1')$ with $2\ell$ bits.\\
 & $(W_1,T_1) \llleftarrow[\rule{0.5cm}{0cm}]{} (W_1',T_1')$ & \\
 {\bf If} $T_1 \neq \Raz(Y_{13}, Z_1)$ {\em reject}.&&\\

&\fbox{{\bf Phases $2$..$L$}}&\\
{\bf For} $i=2$ {\bf to} $L$&  &{\bf For} $i=2$ {\bf to} $L$\\
~~~Sample random $(Y_{i2}, Y_{i3})$.& & \\
~~~$V_{i-1} = \Ext_2(X, W_{i-1})$ with $2\ell$ bits. & & \\
~~~$R_i=\laext(X, Y_{i2})$. \\ 
~~~$Z_i=\lamac_{R_i}(M_i)$. \\
 & $(V_{i-1}, M_i, Y_{i2}, Y_{i3}) \llrightarrow[\rule{0.5cm}{0cm}]{} (V_{i-1}', M_i', Y_{i2}', Y_{i3}')$ & \\
& & ~~~{\bf If} $V_{i-1}'\neq \Ext_2(X, W_{i-1}')$ {\em reject}.\\
& &~~~Sample random $W_i'$ with $d_2$ bits.\\
& & ~~~$R_i'=\laext(X, Y_{i2}')$.\\
& & ~~~$Z_i'=\lamac_{R_i'}(M_i')$.\\
& & ~~~$T_i'= \Raz(Y'_{i3}, Z_i')$ with $2\ell$ bits.\\
 & $(W_i,T_i) \llleftarrow[\rule{0.5cm}{0cm}]{} (W_i',T_i')$ & \\
~~~{\bf If} $T_i \neq \Raz(Y_{i3}, Z_i)$ {\em reject}.\\

{\bf EndFor}&&{\bf EndFor}\\
&\fbox{{\bf Phase $L+1$}}&\\
&&$M'=M_1' \circ \cdots \circ M_L'$. \\
&& {\bf If} $M' \neq \Edit(Y')$ {\em reject}. \\
 $R= \Ext_1(X, Y)$ with $4d_1/\rho$ bits.&& $R'= \Ext_1(X, Y')$ with $4d_1/\rho$ bits.\\
 $V_L = \Ext_2(X, W_L)$ with $2\ell$ bits. & & \\
  & $(V_L) \llrightarrow[\rule{0.5cm}{0cm}]{} (V_L')$ & \\
  & & {\bf If} $V_L'\neq \Ext_2(X, W_L')$ {\em reject}.\\
&& Sample random $W'$ with $d_1$ bits.\\
&& $T' = \mac_{R'}(W')$.\\
&&  Set final $R_B = \Ext_1(X, W')$.\\
 & $(W,T) \llleftarrow[\rule{0.5cm}{0cm}]{} (W',T')$ & \\
{\bf If} $T \neq \mac_{R}(W)$ {\em reject}.&&\\
Set final $R_A = \Ext_1(X, W)$.&&\\
\hline
\end{tabular}
\end{small}
\caption{\label{fig:AKA2}
$(2L+2)$-round Privacy Amplification Protocol for $\thinf(X|E) \geq k$.
}
\end{center}
\end{figure}

We now have the following theorem.

\BT \label{thm:Echange}
The probability that Eve can successfully change $Y$ into $Y' \neq Y$ without causing either Alice or Bob to reject is at most $2^{-\Omega(s)}$.
\ET

\begin{thmproof}
We analyze the transcript of the protocol in Eve's view. Normally, Eve should do alternate interactions with Alice and Bob to send the encoded string $M$. However, since Eve is adversarial, she may do several interactions with Alice or Bob before she resumes interaction with the other. If Eve interacts with Alice twice before she interacts with Bob, then this can be viewed as deleting the first block of message that Alice sends. We call this operation ``D".  If Eve interacts with Bob twice before she interacts with Alice, then this can be viewed as inserting a block of message to Bob. We call this operation ``I". If Eve does not do the above two operations but changes some $M_i$ into a different string $M_i'$ and sends it to Bob, then this can be viewed as altering this block of message. We call this operation ``A". 

Now if Eve successfully changes $Y$ into $Y' \neq Y$ without causing either Alice or Bob to reject, then she must also successfully changes $M=\Edit(Y)$ to $M'=\Edit(Y')$ without causing either Alice or Bob to reject, by a series of $(D, I, A)$ operations. During these operations, we say that at some point Eve has to answer a challenge if Eve has to correctly guess the value of a string that is (close to) uniform even conditioned on the fixing of all transcripts up to this time. We now have the following lemma.

\BL \label{lem:plm1}
For all $(D, I, A)$ operations, except $A$ operations that are immediately followed by $I$ operations, Eve has to answer a challenge.
\EL

\begin{proof}[Proof of the lemma.]
We shall be imprecise about the numbers here. The exact numbers will appear in our next lemma. Note that in the whole protocol the total size of the messages that contain information about $X$ (the $(V, V')$s and $(T, T')$s) is at most $L (8\ell)=d_1/(\rho d_2) \cdot (8\ell) < d_1/\rho$ and $k> d_1/\rho+2s+15d_2^2$. Thus at any time even if conditioned on the fixing of the transcript, $X$ still has a lot of entropy. 

Now if Eve performs a $D$ operation after Alice sends out $(V_{i-1}, M_i, Y_{i2}, Y_{i3})$, then by definition Eve is going to interact with Alice again without interacting with Bob. However Alice is not going to do anything until she receives a response $T_i$ from Bob and checks that $T_i=\Raz(Y_{i3}, Z_i)$. By the same analysis in \theoremref{thm:protocol1}, even if conditioned on the transcript, $\Raz(Y_{i3}, Z_i)$ is close to uniform. Thus Eve has to answer a challenge.

If Eve performs an $I$ operation after Bob sends out $(W_i',T_i')$, then by definition Eve is going to interact with Bob again without interacting with Alice. However Bob is not going to do anything until he receives a response $V_i'$ from Alice and checks that $V_i' = \Ext_2(X, W_i')$. Since conditioned on the transcript $X$ has a lot of entropy and $W_i'$ is uniform and independent of the transcript and $X$, we have that $\Ext_2(X, W_i')$ is close to uniform. Thus Eve has to answer a challenge.

If Eve performs an $A$ operation that is not followed by an $I$ operation, then by definition Eve alters an message $M_i$ to $M_i'$, sends it to Bob and next she is going to interact with Alice (otherwise Eve is going to perform an $I$ operation). Conditioned on the fixing of the transcript before Alice sends out $(V_{i-1}, M_i, Y_{i2}, Y_{i3})$, this is exactly the 2-round protocol as in \theoremref{thm:protocol1}. Since conditioned on the transcript $X$ has a lot of entropy and $M_i \neq M_i'$, by the same analysis in \theoremref{thm:protocol1}, even if further conditioned on the transcript of these two rounds, $\Raz(Y_{i3}, Z_i)$ is close to uniform. Thus Eve has to answer a challenge.

We note that if Eve performs an $A$ operation followed by an $I$ operation, then the above argument may not work (Eve may not have to answer a challenge for the $A$ operation), because the subsequent messages sent out by Bob induced by the $I$ operation may give additional information about $\Raz(Y_{i3}, Z_i)$.
\end{proof}

Our next lemma bounds the probability that Eve successfully answers a challenge.

\BL \label{lem:plm2}
For any $i \in \N$, let $H_i$ stand for the event that Eve successfully answers the $i$'th challenge and $E_i=\cap_{j=1}^{i} H_j$ stand for the event that Eve successfully answers all the challenges up to the $i$'th challenge. Then if $\Pr[E_i] > 2^{-s}$, we have 
\[\Pr[H_{i+1}|E_i] < 2^{-\ell}.\]
\EL

\begin{proof}[Proof of the lemma.]
Note that in the whole protocol the total size of the messages that contain information about $X$ (the $V$s and $T$s) is at most $L (8\ell)=d_1/(\rho d_2) \cdot (8\ell) < d_1/\rho$ and $k> d_1/\rho+2s+15d_2^2$. Thus by \lemmaref{lem:condition}, at any time, with probability $1-2^{-2s}$ over the fixing of the previous transcript, $X$ has min-entropy at least $k-d_1/\rho-2s> 15d_2^2$. 

Now we fix the transcript up to the time before Eve answers the $i+1$'th challenge. The transcript thus determines if Eve successfully answers all previous $i$ challenges. Now consider the transcripts that are in $E_i$. If $\Pr[E_i] > 2^{-s}$, we have that conditioned on $E_i$, with probability at least $1-{2^{-2s}}/\Pr[E_i]> 1-2^{-s}$ over the fixing of the transcript, $X$ has min-entropy at least $15d_2^2$. 

Now assume $X$ indeed has min-entropy at least $15d_2^2$. If for the $i+1$'th challenge, Eve performs a $D$ operation or an $A$ operation not followed by an $I$ operation, then by the same analysis in \theoremref{thm:protocol1}, $\Pr[H_{i+1}] \leq O(2^{-2\ell})$. If Eve performs an $I$ operation, then by \theoremref{thm:optext}, we have $(\Ext_2(X, W_i'), W_i') \approx_{2^{-10\ell}} (U_{2\ell}, W_i')$. Thus we have $\Pr[H_{i+1}] \leq 2^{-2\ell}+2^{-10\ell}=O(2^{-2\ell})$. Adding back the error $2^{-s}$, we have

\[\Pr[H_{i+1}|E_i] \leq O(2^{-2\ell})+2^{-s} < 2^{-\ell}.\]
\end{proof}

Our last lemma bounds the number of challenges that Eve has to answer.

\BL \label{lem:plm3}
If Eve successfully changes $Y$ into $Y' \neq Y$ without causing either Alice or Bob to reject, then she successfully answers at least $2eL/3$ challenges,  where $e$ is the constant in \theoremref{thm:editcode}. 
\EL

\begin{proof}[Proof of the lemma.]
If Eve successfully changes $Y$ into $Y' \neq Y$, then she also successfully changes $M=\Edit(Y)$ to $M'=\Edit(Y')$. Let $a$ be the number of $D$ operations Eve performs, $b$ be the number of $I$ operations Eve performs and $c$ be the number of $A$ operations Eve performs. Since an operation on a block with size $d_2$ is at most $d_2$ operations on the bits, by the property of the edit distance code, we have 

\[(a+b+c)d_2 \geq ed_2L.\]

Thus

\[(a+b+c) \geq eL.\]

By \lemmaref{lem:plm1}, only $A$ operations that are immediately followed by $I$ operations may not cause Eve to answer a challenge. We now bound the number of such $A$ operations.

Let $d$ stand for the number of $A$ operations that are immediately followed by $I$ operations. Thus $d \leq c$ and $d \leq b$. Note that the length of the codeword is fixed, thus we must have $a=b$ and therefore $d \leq a$. Thus we have

\[d \leq (a+b+c)/3.\]

Therefore the number of challenges that Eve successfully answers is at least 

\[a+b+c-d \geq 2(a+b+c)/3 \geq  2eL/3.\]
\end{proof}

Now let $q \geq 2eL/3$ be the number of challenges that Eve successfully answers. Then the probability that this happens is (let $E_0$ be the event that is always true)

\[\Pr[E_q]=\Pi_{j=1}^{q} \Pr[H_j|E_{j-1}].\]

Now if for some $1 \leq j \leq q-1$ we have $\Pr[E_j] \leq 2^{-s}$, then we are already done because $\Pr[E_q] \leq \Pr[E_j] \leq 2^{-s}$. Otherwise by \lemmaref{lem:plm2} we must have that for any  $1 \leq j \leq q$, $\Pr[H_j|E_{j-1}] < 2^{-\ell}$. Thus we have 

\[\Pr[E_q]=\Pi_{j=1}^{q} \Pr[H_j|E_{j-1}] < (2^{-\ell})^q \leq (2^{-\ell})^{2eL/3}=2^{-\Omega(d_1)}=2^{-\Omega(s)}.\]
\end{thmproof}

We now have the following theorem.

\BT
There exists a constant $C>1$ such that for any $k, n \in \N$ with $k \geq \log^4 n$ and any $\e>0$ with $k \geq C(\log(1/\e))$ there exists an explicit $O((\log n+\log(1/\e))/\sqrt{k})$ round privacy amplification protocol for $(n, k)$ sources with security parameter $\log(1/\e)$, entropy loss $O(\log n+\log(1/\e))$ and communication complexity $O((\log n+\log(1/\e))\sqrt{k})$.
\ET

\begin{thmproof}
Without loss of generality we assume that $\e < 2^{-\Omega(\sqrt{k})}$, otherwise we can use the 2-round protocol in \theoremref{thm:protocol1}. Now we show that the protocol in Figure~\ref{fig:AKA2} is such a protocol.

First, if Eve is passive then with probability $1$ Alice and Bob agrees on the random string $W=W'$. Note that the random variables that contain information about $X$ which are used to authenticate $Y$ are $\{V_i, V_i', T_i, T_i'\}$, and the total size of these random variables is at most $L (8\ell)=d_1/(\rho d_2) \cdot (8\ell) < d_1/\rho$. Note that the random variable used to authenticate $W'$ is $R=R'$, which has size at most $4d_1/\rho$. Thus the total size of the random variables in the transcript that contain information about $X$ is at most $5d_1/\rho=O(s)$. Thus we have that conditioned on the fixing of $(Y, \{M_i, V_i, V_i', T_i, T_i', Y_{i2}, Y_{i3}, W_i\}, R)$, the average conditional min-entropy of $X$ is $k-O(s)$, and $W$ is independent of $X$. Thus by \theoremref{thm:optext} we have that $R_A=R_B$ is $2^{-s}$-close to being uniform conditioned on all the transcript, and the entropy loss is $O(s)$.   

Next, if Eve is active and want to make $R_A \neq R_B$, then she has to change $W'$ into a different $W$. Now we have two cases. If Eve does not change $Y$, then we have $Y=Y'$ and thus by by \theoremref{thm:optext} $R=R'$ and is $2^{-s}$-close to being private and uniform even conditioned on $Y$. Note that conditioned on the fixing of $Y$, $R=R'$ is a deterministic function of $X$ with size $4d_1/\rho$. Since the total size of the random variables in the transcript up till now that contain information about $X$ is at most $d_1/\rho$, by \lemmaref{lem:amentropy} the average conditional min-entropy of $R$ is at least $3d_1/\rho$. Thus, by \theoremref{thm:mac} the probability that Eve can change $W'$ into a different $W$ without causing Alice to reject is at most $\rho/2 \cdot 2^{-d_1/\rho}< 2^{-s}$. In the other case, by theorem~\ref{thm:Echange} the probability that Eve can successfully change $Y$ into $Y' \neq Y$ without causing either party to reject is at most $2^{-\Omega(s)}$.

Finally, if Bob does not reject then he computes his own $R_B=\Ext_1(X, W')$ where $W'$ is a random string sampled from his own random bits. Thus in this case we must have $R_B$ is $2^{-s}$-close to being private and uniform. Thus we must have $\Delta((R_B, E'),(\purify(R_B), E')) \leq 2^{-s}$. Now if Eve is passive then clearly $R_A$ is also $2^{-s}$-close to being private and uniform. If Eve is active and does not change $Y$, then by the above analysis if $W' \neq W$ then Alice rejects with probability $1-2^{-s}$. Now consider the probability that Alice rejects when Eve is active and changes $Y$. Let $A$ stand for the event that Alice rejects in this case, and $B$ stand for the event that Bob rejects in this case. By theorem~\ref{thm:Echange} we have

\[\Pr[B]+\Pr[A|\bar{B}]\Pr[\bar{B}] \geq 1-2^{-\Omega(s)}.\]

Now if Bob rejects, then Bob will not send $(W', T')$ to Alice. Thus in this case for Alice not to reject, Eve has to come up with a string $W$ and the correct tag $T=\mac_{R}(W)$ for Alice. Note that in this case conditioned on the transcript,  the average conditional min-entropy of $R$ is still at least $3d_1/\rho$. Thus by theorem~\ref{thm:mac} the probability that Eve can do this without causing Alice to reject is at most $2^{-s}$. Thus we have

\[\Pr[A|B] \geq 1-2^{-s}.\]

Therefore

\begin{align*}
\Pr[A]&=\Pr[A|B]\Pr[B]+\Pr[A|\bar{B}]\Pr[\bar{B}] \geq (1-2^{-s})\Pr[B]+\Pr[A|\bar{B}]\Pr[\bar{B}] \\ &\geq (1-2^{-s})(\Pr[B]+\Pr[A|\bar{B}]\Pr[\bar{B}]) \geq (1-2^{-s})(1-2^{-\Omega(s)}) \\ &\geq 1-2^{-\Omega(s)}.
\end{align*}

Thus, in the case where Eve is active, Alice rejects with probability $1-2^{-\Omega(s)}$. Therefore we must have $\Delta((R_A, E'),(\purify(R_A), E')) \leq 2^{-\Omega(s)}$. Now by choosing an appropriate $s=O(\log(1/\e))$ we have that $2^{-\Omega(s)} \leq \e$ and the entropy loss is $O(\log n+\log(1/\e))$. The number of rounds is $2(L+1)=O(d_1/d_2)=O(s/\ell)=O((\log n+\log(1/\e))/\sqrt{k})$ and the communication complexity is $O(Ld_2^2)=O(d_1d_2)=O((\log n+\log(1/\e)) \sqrt{k})$.
\end{thmproof}

\section{Non-Malleable Condenser for Linear Min-Entropy}\label{sec:linear}
In this section we give a different non-malleable condenser for $(n, k)$ sources with $k =\delta n$ for any constant $0<\delta<1$. This construction has the advantage that the security parameter can achieve up to $\Omega(k)$ instead of $\Omega(\sqrt{k})$. The basic ingredient is a modified alternating extraction protocol borrowed from \cite{Li12b}. 

\begin{figure}[htb]
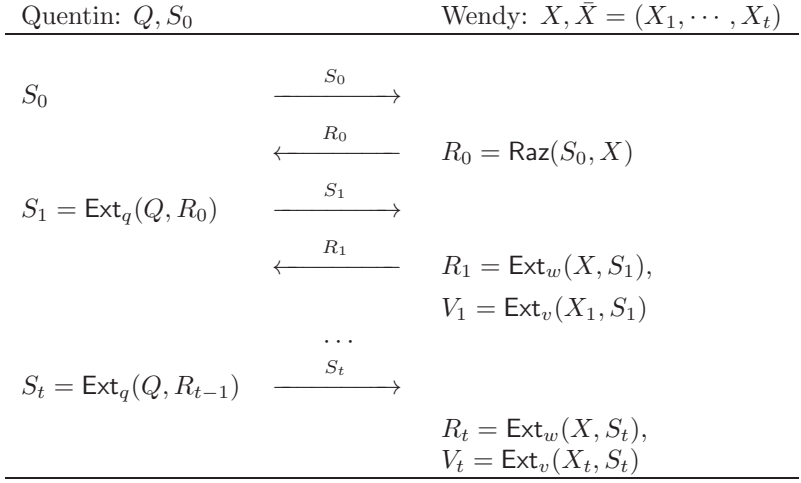

\begin{center}
\begin{small}
\begin{tabular}{l c l}
Quentin:  $Q, S_0$ & &Wendy: $X, \bar{X}=(X_1, \cdots, X_t)$ \\

\hline\\
$S_0$ & $\llrightarrow[\rule{1.5cm}{0cm}]{S_0}{} $ & \\
& $\llleftarrow[\rule{1.5cm}{0cm}]{R_0}{} $ & $R_0=\Raz(S_0, X)$ \\
$S_1=\Ext_q(Q, R_0)$ & $\llrightarrow[\rule{1.5cm}{0cm}]{S_1}{} $ & \\
& $\llleftarrow[\rule{1.5cm}{0cm}]{R_1}{} $ & $R_1=\Ext_w(X, S_1)$, \\
& & $V_1=\Ext_v(X_1, S_1)$ \\
& $\cdots$ & \\
$S_t=\Ext_q(Q, R_{t-1})$ & $\llrightarrow[\rule{1.5cm}{0cm}]{S_t}{} $ & \\
& & $R_t=\Ext_w(X, S_t)$, \\
& & $V_t=\Ext_v(X_t, S_t)$ \\
\hline
\end{tabular}
\end{small}
\caption{\label{fig:altext}
Alternating Extraction.
}
\end{center}
\end{figure}

\textbf{Alternating Extraction.} \cite{Li12b} Assume that we have two parties, Quentin and Wendy. Quentin has a source $Q$ and a source $S_0$ with entropy rate $>1/2$. Wendy has a source $X$ and a source $\bar{X}=(X_1\circ \cdots \circ X_t)$. Suppose that $(Q, S_0)$ is kept secret from Wendy and $(X, \bar{X})$ is kept secret from Quentin.  Let $s, d$ be two parameters for the protocol. Let $\Ext_q$, $\Ext_w$, $\Ext_v$ be seeded extractors as in \theoremref{thm:optext}. Let $\Raz$ be the two-source extractor in \theoremref{thm:razweakseed}. The \emph{alternating extraction protocol} is an interactive process between Quentin and Wendy that runs in $t+1$ steps. 

In the $0$'th step, Quentin sends $S_0$ to Wendy, Wendy computes $R_0=\Raz(S_0, X)$ and replies $R_0$ to Quentin, Quentin then computes $S_1=\Ext_q(Q, R_0)$. In this step $R_0, S_1$ each outputs $d$ bits. In the first step, Quentin sends $S_1$ to Wendy, Wendy computes $V_1=\Ext_v(X_1, S_1)$ and $R_1=\Ext_w(X, S_1)$. She sends $R_1$ to Quentin and Quentin computes $S_2=\Ext_q(Q, R_1)$. In this step $V_1$ outputs $2^{t-1}s$ bits, and $R_1, S_2$ each outputs $d$ bits. In each subsequent step $i$, Quentin sends $S_i$ to Wendy, Wendy computes $V_i=\Ext_v(X_i, S_i)$ and $R_i=\Ext_w(X, S_i)$. She replies $R_i$ to Quentin and Quentin computes $S_{i+1}=\Ext_q(Q, R_i)$. In step $i$, $V_i$ outputs $2^{t-i}s$ bits, and $R_i, S_{i+1}$ each outputs $d$ bits. Thus, the process produces the following sequence: 
\begin{align*}
&S_0, R_0=\Raz(S_0, X),  S_1=\Ext_q(Q, R_0), \\ &V_1=\Ext_v(X_1, S_1), R_1=\Ext_w(X, S_1), \cdots,  \\ &S_t=\Ext_q(Q, R_{t-1}), V_t=\Ext_v(X_t, S_t), R_t=\Ext_w(X, S_t).
\end{align*}

\textbf{Look-Ahead Extractor.} Let $Y=(Q, S_0)$ be a seed, the look-ahead extractor is defined as 
\[\laext((X, \bar{X}),  Y) \eqdef V_1, \cdots, V_t.\]

The following lemma is proved in \cite{Li12b}.

\BL \label{lem:altext2}\cite{Li12b}
In the alternating extraction protocol, assume that $X$ has $n$ bits and $Q, X_i$ each has at most $n$ bits. Let $d=O(\log n+s)> s$ be the number of random bits needed in \theoremref{thm:optext} to achieve error $2^{-s}$. Let $\bar{X}'=(X_1' \circ \cdots \circ X_t')$ be another distribution on the same support of $\bar{X}$ and $(Q', S_0')$ be another distribution on the same support of $(Q, S_0)$ such that $(Q, S_0, Q', S_0')$ is independent of $(X, \bar{X}, \bar{X}')$. Assume that $X$ has min-entropy at least $2^t(4s)+2td$, $Q$ has min-entropy at least $4td+60d+6s$ and $S_0$ is a $(30d+3s, 29d+2s)$ source. 

Now run the alternating extraction protocol with $(X, \bar{X}')$ and $(Q', S_0')$ where in each step we obtain $S_i', R_i', V_i'$.  For any $i, 0 \leq i \leq t$, let $View_i=(S_0, \cdots, S_{i}, R_0, \cdots, R_{i}, V_1, \cdots, V_i)$ and let $View_i'=(S_0', \cdots, S_{i}', R_0', \cdots, R_{i}', V_1', \cdots, V_i')$. Then if for some $j \leq t$, $\bar{X}_j$ has min-entropy at least $2^t(3s)+2td$, we have

\[(V_j, S_j, S'_j, View_{j-1}, View'_{j-1}, Q, Q') \approx_{O(t2^{-s})} (U_{2^{t-j}s}, S_j, S'_j, View_{j-1}, View'_{j-1}, Q, Q').\]
\EL

We now describe our non-malleable condenser.

\begin{algorithmsub}
{$\nmC(x, y)$}
{$\ell$--an integer parameter. $x$ --- a sample from an $(n,k)$-source with $k \geq \delta n$. $y$--an independent random seed with $y=(y_1, y_2)$. }
{$z$ --- an $m$ bit string.} {
Let $d=O(\log n+\ell)$ be the length of a seed that can achieve error $2^{-5\ell}$ for both the non-malleable extractor in \theoremref{thm:enmext} and the strong extractor in \theoremref{thm:optext}. 

Let $\Cond:\zo^n \rightarrow (\zo^{n'})^C$ be a rate-$(\delta \to 0.9, 2^{-2\ell})$\\-somewhere-condenser as in \theoremref{thm:swcondenser}, where $C=\poly(1/\delta)$, $n'=\poly(\delta)n$.

Let $\nmExt:\bits^{n'}\times \bits^{d}\rightarrow \bits^{m'}$ be a $(0.8n',2^{-2\ell})$\\-non-malleable extractor as in \theoremref{thm:enmext} with output length $m' = 6\cdot 2^C\ell$.

Let $y_1$ be a random string with $d$ bits, $y_2$ be a random string with $d'=4Cd+61d+14\ell$ bits. 

Let $\nmExt_2:\bits^{2^C(10\ell)}\times \bits^{d'}\rightarrow \bits^{2^C (4\ell)}$ be a $(2^C(10\ell),2^{-4\ell})$-non-malleable extractor as in \theoremref{thm:enmext}.

Let $\laext$ be the look-ahead extractor defined above, with parameters $(2\ell, d)$ and using $q=y_2$ and $s_0$ is the first $30d+6\ell$ bits of $y_2$.

} {alg:nmc2}

\begin{enumerate}

\item Compute $(x_1,\ldots x_C) = \Cond(x)$. 

\item Compute $w=\Ext(x, y_1)$ with output size $2^C(10\ell)$. 

\item  Compute $\bar{x}=(\bar{x}_1, \ldots, \bar{x}_C)$ where $\bar{x}_i=\nmExt(x_i, y_1)$.

\item Compute $v=(v_1, \ldots, v_C)=\laext((x, \bar{x}), y_2)$.

\item Output $z=(\nm_2(w, y_2), v)$ such that $\nm_2(w, y_2)$ has size $2^C (4\ell)$. 
\end{enumerate}
\end{algorithmsub}

We have the following theorem.

\BT \label{thm:nmc2}
For any constant $0<\delta<1$ and $k=\delta n$ there exists a constant $C_1=2^{\poly(1/\delta)}$ such that given any $0<s \leq k/C_1$, the above construction is a $(k, s, 2^{-s})$-non-malleable condenser with seed length $\poly(1/\delta)(\log n+s)$.
\ET

\begin{thmproof}
Let $\adv$ be any (deterministic) function such that $\forall y \in \Supp(Y), \adv(y) \neq y$. We will show that for most $y$, with high probability over the fixing of $\nmC(X, \adv(y))$, $\nmC(X, y)$ is still close to having min-entropy at least $\ell$. Let $Y'=\adv(Y)$. Thus $Y' \neq Y$. In the following analysis we will use letters with prime to denote the corresponding random variables produced with $Y'$ instead of $Y$. Let $H=\nm_2(W, Y_2)$. Thus $Z=(H, V)$. We have the following two cases.

\textbf{Case 1:} $Y_1 = Y'_1$. In this case, since $Y' \neq Y$, we must have that $Y_2 \neq Y'_2$. Now by \theoremref{thm:optext} we have that 

\[(W, Y_1) \approx_{2^{-{5\ell}}} (U, Y_1).\]

Therefore, we can now fix $Y_1$ (and thus $Y'_1$), and with probability $1-2^{-\ell}$ over this fixing, $W$ is $2^{-4\ell}$-close to uniform. Moreover, after this fixing $W$ is a deterministic function of $X$ and thus is independent of $Y_2$. Note also that after this fixing, $Y'_2$ is a deterministic function of $Y_2$. Thus by \theoremref{thm:enmext} we have that 

\[(H, H', Y_2, Y'_2) \approx_{O(2^{-4\ell})} (U_{2^C (4\ell)}, H', Y_2, Y'_2).\] 

Therefore, we can now further fix $Y_2$ (and thus $Y'_2$) and with probability at least $1-O(2^{-\ell})$ over this fixing, $(H, H')$ is $2^{-3\ell}$-close to $(U_{2^C (4\ell)}, H')$. Thus we can further fix $H'$, and with probability at least $1-2^{-\ell}$ over this fixing, $H$ is $2^{-2\ell}$-close to uniform. Now note that $H$ has size $2^C (4\ell)$ and $V'$ has size at most $2^C (2\ell)$. Thus by \lemmaref{lem:econdition}, we can further fix $V'$, and with probability at least $1-2 \cdot 2^{-\ell}$ over this fixing, $V_1$ is $2^{\ell}$-close to having min-entropy at least $2^C (4\ell)-2^C (2\ell) -\ell  \geq 3\ell$.

Thus in this case we have shown that, with probability $1-O(2^{-\ell})$ over the fixing of $Y$, with probability $1-O(2^{-\ell})$ over the fixing of $Z'$, $Z$ is $2^{-\ell}$-close to having min-entropy at least $3\ell> 2\ell$.

\textbf{Case 2:} $Y_1 \neq Y'_1$. In this case, first note that by \theoremref{thm:swcondenser}, $\Cond(X)=(X_1,\ldots X_C)$ is $2^{-\ell}$-close to a somewhere rate-$0.9$-source with $C$ rows, and each row has length $\Omega(n)$. In the following we will simply treat it as a somewhere rate-$0.9$-source, since this only adds $2^{-\ell}$ to the error. We assume that $X_g, 1 \leq g \leq C$ is a rate $0.9$-source \footnote{In general a somewhere rate-$0.9$-source is a convex combination of elementary somewhere rate-$0.9$-sources, but without loss of generality we can assume it is an elementary somewhere rate-$0.9$-source.}. 

Now since the adversary changes $Y_1$ to $Y'_1 \neq Y_1$, by \theoremref{thm:enmext} we have that 

\[(\bar{X}_g, \bar{X}'_g, Y_1) \approx_{2^{-2\ell}} (U_{m'},  \bar{X}'_g, Y_1).\]

As the first step for the following analysis, we now fix $Y_1$, $Y'_1$ and $W'=\Ext(X, Y_1'), \bar{X}'_g$. Note that $Y'_1$ is a deterministic function of $(Y_1, Y_2)$, and after fixing $Y'_1$, $(W', \bar{X}'_g)$ is a deterministic function of $X$. Thus by \lemmaref{lem:amentropy} we have the following claim.

\BCM \label{clm:condition0}
After the fixings of $(Y_1, Y'_1, W', \bar{X}'_g)$, $\bar{X}_g$ is a deterministic function of $X$ and is $2^{-\ell}$ close to a source with average conditional min-entropy $m'-2^C(4\ell)$. 
\ECM

Note that by \lemmaref{lem:amentropy}, after this fixing, the average conditional min-entropy of $X$ is at least $k-m'-2^C(4\ell)$ and $m'=\poly(\delta)n$. Thus for a sufficiently small $\ell=\Omega(k)$ we can ensure that $k-m'-2^C(4\ell) \geq 2^C(8\ell)+2Cd$ and $m'-2^C(4\ell) \geq 2^C(6\ell)+2Cd$. Since $Y_1$ is independent of $Y_2$ and $Y'_1$ is a deterministic function of $(Y_1, Y_2)$, by \lemmaref{lem:condition} we have that with probability $1-2 \cdot 2^{-2\ell}$ over this fixing, $Q=Y_2$ is a source with min-entropy at least $4Cd+60d+12\ell$ and $S_0$ is a source with min-entropy $29d+4\ell$. Now by \lemmaref{lem:altext2} (and note that $s=2\ell$) we have that 

\[(V_g, S_g, S'_g, View_{g-1}, View'_{g-1}, Y_2, Y_2') \approx_{O(C2^{-2\ell})} (U_{2^{C-g}(2\ell)}, S_g, S'_g, View_{g-1}, View'_{g-1}, Y_2, Y_2').\]

Adding back all the error, and noticing that we have fixed $(Y_1, Y'_1, W', \bar{X}'_g)$ before, we have

\begin{align*}
&(V_g, S_g, S'_g, View_{g-1}, View'_{g-1}, W', \bar{X}'_g, Y_1, Y_1', Y_2, Y_2') \\ \approx_{O(C2^{-2\ell})} &(U_{2^{C-g}(2\ell)}, S_g, S'_g, View_{g-1}, View'_{g-1}, W', \bar{X}'_g, Y_1, Y_1', Y_2, Y_2').
\end{align*}

Note that $V'_g=\Ext_v(\bar{X}'_g, S'_g)$ and $H'=\nm_2(W', Y_2')$. Thus we have that 

\[(V_g, View'_{g-1}, H', V'_g, Y) \approx_{O(C2^{-2\ell})} (U_{2^{C-g}(2\ell)}, View'_{g-1}, H', V'_g, Y).\]

This implies that 

\[(V_g, H', V'_1, \cdots, V'_g, Y) \approx_{O(C2^{-2\ell})} (U_{2^{C-g}(2\ell)}, H', V'_1, \cdots, V'_g, Y).\]

Thus we have that with probability $1-O(C2^{-\ell/2})$ over the fixing of $Y$, 

\[(V_g, H', V'_1, \cdots, V'_g) \approx_{2^{-3\ell/2}} (U_{2^{C-g}(2\ell)}, H', V'_1, \cdots, V'_g).\] 

Thus, with probability $1-2^{-\ell/2}$ over the further fixing of $(H', V'_1, \cdots, V'_g)$, we have $V_g \approx_{2^{-\ell}} U_{2^{C-g}(2\ell)}$. Now note the size of $(V'_{g+1}, \cdots, V'_C)$ is at most $\sum_{i=g+1}^{C}2^{C-i}(2\ell) = 2^{C-g}(2\ell)-2\ell$, and that $V_g$ has size $2^{C-g}(2\ell)$. Thus by \lemmaref{lem:econdition}, with probability $1-2 \cdot 2^{-\ell/2}$ over the further fixing of $(V'_{g+1}, \cdots, V'_C)$, we have that $V_g$ is $2^{-\ell/2}$-close to a source with min-entropy $2\ell-\ell/2>\ell$. Since $V'=(V'_1, \cdots, V'_g, V'_{g+1}, \cdots, V'_C)$ and $Z'=(H', V')$, altogether in this case we have that with probability $1-O(C2^{-\ell/2})$ over the fixing of $Y$, with probability $1-2^{-\ell/2}$ over the further fixing of $Z'$, $V_g$ is $2^{-\ell/2}$-close to a source with min-entropy $>\ell$. Thus $Z$ is also $2^{-\ell/2}$-close to a source with min-entropy $>\ell$. 

Combining \textbf{Case 1} and \textbf{Case 2}, and notice that the fraction of ``bad seeds" that an adversary can achieve is at most the sum of the fraction of bad seeds in both cases. Thus we have that with probability $1-O(C2^{-\ell/2})$ over the fixing of $Y$, with probability $1-2^{-\ell/2}$ over the further fixing of $Z'$, $Z$ is $2^{-\ell/2}$-close to a source with min-entropy $>\ell$. by choosing an appropriate $\ell=O(s)$ we have that the construction is a $(k, s, 2^{-s})$-non-malleable condenser with seed length $O(Cd)=\poly(1/\delta)(\log n+s)$. 
\end{thmproof}

Combining this theorem with \theoremref{thm:nmcext}, we get the following theorem.

\BT
There exists an absolute constant $C_0>1$ such that for any constant $0< \delta <1$ and $k =\delta n$ there exists a constant $C_1=2^{\poly(1/\delta)}$ such that given any $\e>0$ with $C_1\log(1/\e) \leq k$, there exists an explicit 2-round privacy amplification protocol for $(n, k)$ sources with security parameter $\log(1/\e)$, entropy loss $C_0(\log n+\log (1/\e))$ and communication complexity $\poly(1/\delta)(\log n+\log(1/\e))$.
\ET

\section{Conclusions and Open Problems}\label{sec:conc}
In this paper we construct explicit non-malleable condensers for arbitrary min-entropy, and use them to give an explicit 2-round privacy amplification protocol with optimal entropy loss for arbitrary min-entropy $k$, with security parameter up to $s=\Omega(\sqrt{k})$. This is the first explicit protocol that simultaneously achieves optimal parameters in both round complexity and entropy loss, for arbitrary min-entropy.

We then generalize this result to give a privacy amplification protocol that runs in $O(s/\sqrt{k})$ rounds and achieves optimal entropy loss for arbitrary min-entropy $k$, with security parameter up to $s=\Omega(k)$. This significantly improves the protocol in \cite{ckor}. In the special case where $k=\delta n$ for some constant $\delta>0$, we give better non-malleable condensers and a 2-round privacy amplification protocol with optimal entropy loss for security parameter up to $s=\Omega(k)$, which improves the entropy loss and communication complexity of the 2-round protocol in \cite{Li12b}. 

Some open problems include constructing better non-malleable extractors or non-malleable condensers, and to construct optimal privacy amplification protocols for security parameter bigger than $\sqrt{k}$. Another interesting problem is to find other applications of non-malleable extractors or non-malleable condensers.
\bibliographystyle{alpha}

\bibliography{refs}

\end{document}